\documentclass[preprint,aps,amsmath,superscriptaddress,nofootinbib,tightenlines,showpacs]{revtex4}
\usepackage{amsmath}
\usepackage{amsfonts}
\usepackage{amssymb}
\usepackage{latexsym}
\usepackage{graphicx}
\usepackage{bm}
\usepackage{epsfig}
\usepackage[english]{babel}

\def\sign{\mbox{sign}}

\begin{document}


\title{Curvature and Entropy Perturbations in Generalized Gravity}
\author{Xiangdong Ji\footnote{Electronic address: xji@physics.umd.edu}}
\affiliation{Center for High-Energy Physics, Peking University,\\
Beijing 100871, China\\} \affiliation{Maryland Center for
Fundamental Physics and Department of Physics, University of
Maryland,\\ College Park, Maryland 20742, USA\\ \vspace{0.2cm}}
\author{Tower Wang\footnote{Electronic address: wangtao218@pku.edu.cn}}
\affiliation{Center for High-Energy Physics, Peking University,\\
Beijing 100871, China\\}
\date{\today\\ \vspace{1cm}}
\begin{abstract}
We investigate the cosmological perturbations in generalized
gravity, where the Ricci scalar and a scalar field are non-minimally
coupled via an arbitrary function $f(\varphi,R)$. In the
Friedmann-Lema\^{i}tre-Robertson-Walker (FLRW) background, by
studying the linear perturbation theory, we separate the scalar type
perturbations into the curvature perturbation and the entropy
perturbation, whose evolution equations are derived. Then we apply
this framework to inflation. We consider the generalized slow-roll
conditions and the quantization initial condition. Under these
conditions, two special examples are studied analytically. One
example is the case with no entropy perturbation. The other example
is a model with the entropy perturbation large initially but
decaying significantly after crossing the horizon.
\end{abstract}

\pacs{98.80.Cq, 04.50.Kd}

\maketitle

\newpage



\section{Introduction}\label{sect-intro}
In the past three decades, huge progress has been made on our
understanding of the early universe, both theoretically and
observationally. This is implemented by the inflation theory
\cite{Guth:1980zm,Linde:1981mu,Albrecht:1982wi} merging the general
relativity and quantum field theory in an elegant way. On the one
hand, inflation theory has naturally explained the initial condition
of big bang cosmology. On the other hand, it also makes quantitative
predictions which can be tested by precise observational data
\cite{AdelmanMcCarthy:2007wh,Komatsu:2008hk,:2006uk}.

So far the prevail inflation models are based on the Einstein
gravity coupled minimally to a scalar field (or more scalar fields)
\cite{Mukhanov:1990me,Riotto:2002yw,Polarski:1992dq,Polarski:1994rz,Polarski:1995zn,Langlois:1999dw,Gordon:2000hv,Wands:2002bn}.
Whereas considerable investigations were also performed on models of
modified gravity with higher derivative corrections or non-minimal
coupling. Most of them can be classified into two categories:
\begin{itemize}
\item $f(R)$ models without a scalar field \cite{Starobinsky:1980te,Starobinsky:1983sov};
\item $F(\varphi)R$ scalar-tensor theory.
\end{itemize}
Each of them has only one degree of freedom. This is clear if one
takes a conformal transformation as done by
\cite{Teyssandier:1983zz,Maeda:1988ab,Wands:1993uu}.

Our intention in this paper is to deal with a general type of model,
namely $f(\varphi,R)$ gravity, which unifies and generalizes the
above relatively simpler models. The action of this model is of the
form\footnote{Throughout this paper, we employ the reduced Planck
mass $M_p=1/\sqrt{8\pi G}$ and set $\hbar=c=1$.}
\begin{equation}\label{action}
S=\int d^4x\sqrt{-g}\left[\frac{1}{2}f(\varphi,R)-\frac{1}{2}g^{\alpha\beta}\partial_{\alpha}\varphi\partial_{\beta}\varphi-V(\varphi)\right].
\end{equation}
Here the $f(\varphi,R)$ term contains a non-minimal coupling between
the scalar field $\varphi$ and the Ricci scalar $R$. While
$V(\varphi)$ is the potential of the scalar field. In principle
$V(\varphi)$ can be absorbed in $f(\varphi,R)$, but we will keep
them separate.

In contrast with simpler models, the $f(\varphi,R)$ model usually
introduces another degree of freedom. Generally speaking, due to the
new degree of freedom, there is a non-vanishing entropy perturbation
in most models based on $f(\varphi,R)$ gravity. In the present work,
we will distinguish the entropy perturbation from curvature
perturbation, and then study their evolutions.

To make our discussion self-contained and clear in notations,
first of all, we collect some previously known results in section
\ref{sect-review}. Our general result is presented in section
\ref{sect-curen}, where we extract the curvature perturbation and
entropy perturbation as well as their evolution equations.
Applying this formalism to inflation, we study the generalized
slow-roll conditions and the quantized initial condition in
section \ref{sect-slroll}. In section \ref{sect-example}, some
typical examples are studied under the slow-roll approximation.
One example is the case with no entropy perturbation, including
the simpler models with one degree of freedom we mentioned above.
The other example is to add a $g(\varphi)R^2$ correction to
Einstein gravity. Specifically, we study the
$g(\varphi)=\frac{1}{4}\lambda\varphi^2$,
$V(\varphi)=\frac{1}{2}m^2\varphi^2$ model in the limit
$M_p^2/\varphi^2\ll\lambda m^2\varphi^2/M_p^2\ll1$. Ignoring the
coupling of perturbations inside the Hubble horizon, we find the
entropy perturbation is large at horizon-crossing but decays
significantly outside the horizon. Initially the correlation
between the curvature perturbation and the entropy perturbation
has been neglected under our approximation, but at the end of
inflation they become moderately anti-correlated. We summarize and
refer to some open problems in section \ref{sect-sum}. For
reference and as a support to our canonical quantized initial
conditions, in appendix \ref{app-conf} we collect the relevant
results of a two-field model which is conformally equivalent to
the $f(\varphi,R)$ model. In the generalized gravity, since it is
difficult to draw a clear borderline between gravity and the
matter, there is ambiguity in defining curvature perturbation and
entropy perturbation. we present a more traditional (but less
tractable) definition of these perturbations in appendix
\ref{app-perts}. Complementary to section \ref{sect-curen},
details for deriving the evolution equation of entropy
perturbation are relegated to appendix \ref{app-entr}.

\section{Review of Previously Known Results}\label{sect-review}
The cosmological perturbations in $f(\varphi,R)$ gravity has been
studied actively by Hwang and Noh in
\cite{Hwang:1990re,Hwang:1996np,Hwang:1996bc,Hwang:1997uc,Hwang:2005hb}
\emph{etc}. But most of the investigations mainly concentrated on
theories with one degree of freedom, including the $f(R)$ theory and
the scalar-tensor theory. For generalized $f(\varphi,R)$ theory, the
complete evolution equations of the first order perturbations were
obtained in \cite{Hwang:2005hb}, where the background dynamics were
also shown. In this section, to make our discussion self-contained
and clear in notations, we write down these results following the
notations of \cite{Mukhanov:1990me}, except for that the metric
signature is taken to be ($-,+,+,+$). All of the results collected
here can be found in \cite{Hwang:2005hb,Chen:2006wn}. Throughout
this paper, we will focus on the scalar type perturbations, working
in the longitudinal gauge. The tensor type perturbation has been
addressed in \cite{Hwang:1997uc}. There was a discussion of scalar
type perturbations in \cite{Hwang:1996bc}, though, for generalized
gravity, their full evolution equations were obtained in
\cite{Hwang:2005hb}. In this section, we summarize these results in
a self-contained way, and at the same time, set up our convention of
notations. In the subsequent sections, we will take further steps to
get some new results.

The background is described by the
Friedmann-Lema\^{i}tre-Robertson-Walker (FLRW) metric
\begin{equation}
ds^2=-dt^2+a^2(t)d\vec{x}^2=a^2(\tau)(-d\tau^2+d\vec{x}^2),
\end{equation}
where $t$ is the comoving time and $\tau$ is the conformal time,
with respect to which the derivatives will be denoted by a dot
overhead and a superscript prime respectively. Later on, for the
sake of convenience, we will also use a superscript ``$\bullet$'' to
denote the derivative with respect to $t$. Then in terms of the
Hubble parameter $H=\dot{a}/a$, the Ricci scalar\footnote{When
defining the Ricci scalar $R=g^{\mu\nu}R_{\mu\nu}$, we take the
convention of Ricci tensor
$R_{\mu\nu}=\partial_{\lambda}\Gamma^{\lambda}_{\mu\nu}-\partial_{\nu}\Gamma^{\lambda}_{\mu\lambda}+\Gamma^{\lambda}_{\lambda\kappa}\Gamma^{\kappa}_{\mu\nu}-\Gamma^{\lambda}_{\nu\kappa}\Gamma^{\kappa}_{\mu\lambda}$
with the affine connections
$\Gamma^{\lambda}_{\mu\nu}=\frac{1}{2}g^{\lambda\kappa}(\partial_{\mu}g_{\kappa\nu}+\partial_{\nu}g_{\mu\kappa}-\partial_{\kappa}g_{\mu\nu})$.}
can be expressed as $R=6(2H^2+\dot{H})$. Throughout this paper, we
will only deal with the flat universe. The results for closed or
open universe are expected to be similar.

For succinctness let us introduce the notation
$F=\frac{\partial}{\partial R}f(\varphi,R)$. In this paper we will
concentrate on the case with $F>0$, but it is easy to extend our
results to the case $F\leq0$. The fluctuation of the scalar field
$\varphi$ will be denoted by $\delta \varphi$. It is well known that
the perturbations of the metric can be decomposed into three types:
the scalar type, the vector type and the tensor type. In the
scalar-driven inflation, these three types are decoupled with each
other if we only consider two-point correlation functions. Here we
are interested in the linear order scalar type perturbations, so we
can treat them exclusively, not worrying about the vector type and
tensor type perturbations. Considering scalar type perturbations
only, the perturbed metric takes the form
\begin{equation}
ds^2=-(1+2\phi)dt^2+2a\partial_i\mathcal{B}dtdx^i+a^2[(1-2\psi)\delta_{ij}+2\partial_i\partial_j\mathcal{E}]dx^{i}dx^{j}.
\end{equation}
Here $\delta_{ij}$ is the Kronecker delta function. We will mainly
work in the longitudinal gauge, that is, choosing
$\mathcal{B}=\mathcal{E}=0$. If necessary, one can easily recover
all of our results into the gauge-invariant form with the
following dictionary \cite{Mukhanov:1990me,Hwang:2005hb}:
\begin{eqnarray}\label{gauge}
\nonumber \phi&\rightarrow&\phi^{(gi)}=\phi+\frac{[a(\mathcal{B}-\mathcal{E}')]'}{a},\\
\nonumber \psi&\rightarrow&\psi^{(gi)}=\psi-aH(\mathcal{B}-\mathcal{E}'),\\
\nonumber \delta\varphi&\rightarrow&\delta\varphi^{(gi)}=\delta\varphi+a\dot{\varphi}(\mathcal{B}-\mathcal{E}'),\\
\delta F&\rightarrow&\delta F^{(gi)}=\delta
F+a\dot{F}(\mathcal{B}-\mathcal{E}').
\end{eqnarray}

Corresponding to action
(\ref{action}), the variation of $\varphi$ and $g_{\mu\nu}$ gives
\begin{eqnarray}\label{variation}
\nonumber \delta_1S&=&\int d^4x\mathfrak{D}_0+\int d^4x\left[\sqrt{-g}\left(\frac{1}{2}f_{,\varphi}-V_{,\varphi}\right)+\partial_{\nu}(\sqrt{-g}g^{\mu\nu}\partial_{\mu}\varphi)\right]\delta\varphi\\
\nonumber &&+\int d^4x\biggl\{\frac{1}{2}\sqrt{-g}\biggl[-g_{\mu\nu}\left(\frac{1}{2}f-\frac{1}{2}g^{\alpha\beta}\partial_{\alpha}\varphi\partial_{\beta}\varphi-V\right)+FR_{\mu\nu}-\partial_{\mu}\varphi\partial_{\nu}\varphi\\
\nonumber &&+\partial_{\alpha}Fg^{\alpha\beta}\frac{1}{2}(\partial_{\beta}g_{\mu\nu}+\partial_{\mu}g_{\nu\beta}-\partial_{\nu}g_{\mu\beta})-\partial_{\mu}Fg^{\alpha\beta}\frac{1}{2}(\partial_{\alpha}g_{\beta\nu}+\partial_{\beta}g_{\alpha\nu}-\partial_{\nu}g_{\alpha\beta})\biggr]\\
&&+\partial_{\kappa}\left[\frac{1}{4}\sqrt{-g}\partial_{\lambda}F(2g^{\alpha\beta}g^{\lambda\kappa}-g^{\alpha\lambda}g^{\beta\kappa}-g^{\alpha\kappa}g^{\beta\lambda})\right]g_{\alpha\mu}g_{\beta\nu}\biggr\}\delta g^{\mu\nu},
\end{eqnarray}
where the total derivative term
\begin{eqnarray}
\nonumber \mathfrak{D}_0&=&\partial_{\lambda}\left(\frac{1}{2}\sqrt{-g}Fg^{\mu\nu}\delta \Gamma^{\lambda}_{\mu\nu}\right)-\partial_{\nu}\left(\frac{1}{2}\sqrt{-g}Fg^{\mu\nu}\delta \Gamma^{\lambda}_{\mu\lambda}\right)-\partial_{\nu}(\sqrt{-g}g^{\mu\nu}\partial_{\mu}\varphi\delta \varphi)\\
&&+\partial_{\kappa}\left[\frac{1}{4}\sqrt{-g}\partial_{\lambda}F(2g^{\mu\nu}g^{\lambda\kappa}-g^{\mu\lambda}g^{\nu\kappa}-g^{\mu\kappa}g^{\nu\lambda})\delta g_{\mu\nu}\right].
\end{eqnarray}

Then the generalized Friedmann equations are simply
\begin{eqnarray}
\label{Friedmann1}&&\frac{1}{2}\dot{\varphi}^2+V-\frac{1}{2}f+3H^2F+3\dot{H}F-3H\dot{F}=0,\\
\label{Friedmann2}&&\dot{\varphi}^2+2\dot{H}F-H\dot{F}+\ddot{F}=0.
\end{eqnarray}
Equation (\ref{Friedmann2}) proves to be very useful and we will use
it frequently without mention. The non-minimal coupling
$f(\varphi,R)$ brings a new term to the equation of motion for the
scalar field
\begin{equation}\label{eom-varphi}
\ddot{\varphi}+3H\dot{\varphi}-\frac{1}{2}f_{,\varphi}+V_{,\varphi}=0.
\end{equation}

From (\ref{variation}), we get the Einstein equations of linear
perturbations\cite{Hwang:2005hb,Chen:2006wn} in the longitudinal
gauge
\begin{eqnarray}
\label{g00}&&3H(\dot{\psi}+H\phi)-\frac{1}{a^2}\nabla^2\psi=-\frac{1}{2 M_p^{2}}\delta\rho,\\
\label{g0i}&&\dot{\psi}+H\phi=-\frac{1}{2 M_p^{2}}\delta q,\\
\label{gij}&&\psi-\phi=\frac{\delta F}{F},\\
\label{gii}&&\ddot{\psi}+3H(H\phi+\dot{\psi})+H\dot{\phi}+2\dot{H}\phi+\frac{1}{3a^2}\nabla^2(\phi-\psi)=\frac{1}{2M_p^{2}}\delta p,
\end{eqnarray}
in which $\delta\rho$, $\delta q$ and $\delta p$ are defined as
\begin{eqnarray}
\nonumber \delta\rho&=&\frac{M_p^{2}}{F}[\dot{\varphi}\delta\dot{\varphi}+\frac{1}{2}(-f_{,\varphi}+2V_{,\varphi})\delta\varphi-3H\delta\dot{F}+(3\dot{H}+3H^2+\frac{1}{a^2}\nabla^2)\delta F\\
\nonumber &&+(3H\dot{F}-\dot{\varphi}^2)\phi+3\dot{F}(H\phi+\dot{\psi})],\\
\nonumber \partial_i(\delta q)&=&-\frac{M_p^{2}}{F}\partial_i(\dot{\varphi}\delta\varphi+\delta\dot{F}-H\delta F-\dot{F}\phi),\\
\nonumber \delta p&=&\frac{M_p^{2}}{F}[\dot{\varphi}\delta\dot{\varphi}+\frac{1}{2}(f_{,\varphi}-2V_{,\varphi})\delta\varphi+\delta\ddot{F}+2H\delta\dot{F}+(-\dot{H}-3H^2-\frac{2}{3a^2}\nabla^2)\delta F\\
&&-\dot{F}\dot{\phi}-(\dot{\varphi}^2+2\ddot{F}+2H\dot{F})\phi-2\dot{F}(H\phi+\dot{\psi})].
\end{eqnarray}
Equations (\ref{g00}-\ref{gii}) follow respectively from the
$G^0_0$, $G^0_i$, $G^i_j(i\neq j)$, $G^i_i$ components of Einstein
equations. The equation of motion for $\delta \varphi$ gives a
redundant relation.

Using equations (\ref{g0i}) and (\ref{gij}) to cancel $\delta
\varphi$ and $\delta F$ respectively in (\ref{g00}), one obtains
\begin{eqnarray}\label{pert1}
\nonumber&&F(\ddot{\phi}+\ddot{\psi})+(HF+3\dot{F}-\frac{2F\ddot{\varphi}}{\dot{\varphi}})(\dot{\phi}+\dot{\psi})\\
\nonumber &&+[(3H\dot{F}+3\ddot{F})-\frac{2\ddot{\varphi}}{\dot{\varphi}}(HF+2\dot{F})-\frac{F}{a^2}\nabla^2]\phi\\
&&+[(4\dot{H}F+H\dot{F}-\ddot{F})-\frac{2\ddot{\varphi}}{\dot{\varphi}}(HF-\dot{F})-\frac{F}{a^2}\nabla^2]\psi=0.
\end{eqnarray}

Inserting (\ref{g0i}) and (\ref{gij}) into the $G^i_i$ component
equation (\ref{gii}), it will result in a relation automatically
satisfied by the background equation (\ref{eom-varphi}).

Remembering that $\delta F=F_{,R}\delta R+F_{,\varphi}\delta\varphi$
and $\dot{F}=F_{,R}\dot{R}+F_{,\varphi}\dot{\varphi}$, we can write
(\ref{gij}) in another form
\begin{eqnarray}\label{pert2}
\nonumber F(\phi-\psi)&=&2F_{,R}[3\ddot{\psi}+3H\dot{\phi}+12H\dot{\psi}+6(2H^2+\dot{H})\phi+\frac{1}{a^2}\nabla^2(\phi-2\psi)]\\
&&-\frac{F_{,\varphi}}{\dot{\varphi}}[F(\dot{\phi}+\dot{\psi})+(HF+2\dot{F})\phi+(HF-\dot{F})\psi].
\end{eqnarray}
In minimally coupled model $f=M_p^2R$, equation (\ref{pert2})
reduces to the familiar relation $\phi=\psi$.

Comments are needed here. We start with five equations (four
perturbed Einstein equations and one equation of motion for
$\delta\varphi$) of three variables ($\delta\varphi$, $\phi$,
$\psi$), so this system of equations appears to be over-determined.
However, as we have just mentioned, the equation of motion for
$\delta\varphi$ is redundant, which can be derived from perturbed
Einstein equations. In addition, one of the remaining four equations
turns out to be automatically satisfied by background equations.
This is also understandable if one recalls the Bianchi identity.
Consequently, there are three independent equations with three
variable at last. What we have done was just using one of the
equations to cancel $\delta\varphi$, and arriving at two equations
(\ref{pert1}), (\ref{pert2}) with two variables $\phi$, $\psi$.

\section{Curvature Perturbation and Entropy Perturbation}\label{sect-curen}
The perturbation equations (\ref{pert1}) and (\ref{pert2}) were used
in \cite{Hwang:2005hb,Chen:2006wn} to study certain $f(\varphi,R)$
models with only one degree of freedom, where the effects of entropy
perturbation have been neglected all the while. However, in the most
general case, the entropy perturbation may play an important role in
the early stage and then decay or translate into the curvature
perturbation, leaving some observable signatures in fluctuations of
cosmic microwave background (CMB) and dark matter.\footnote{A
mechanism for entropy perturbation decay into curvature perturbation
was realized in the curvaton scenario
\cite{Lyth:2001nq,Lyth:2002my}. In $f(\varphi,R)$ gravity one may
expect a similar story. It is remarkable that by curvaton mechanism
a large non-Gaussianty can be generated
\cite{Lyth:2001nq,Lyth:2002my,Huang:2008ze,Huang:2008rj,Huang:2008bg}.}
Disregarding the entropy perturbation would obscure many interesting
phenomenological predictions, such as the residual entropy
perturbation and the large non-Gaussianity. From this point of view,
for generalized $f(\varphi,R)$ gravity, it is important to take the
new degree of freedom into consideration and study the entropy
perturbation seriously.

In this section, we will rearrange the perturbation equations
(\ref{pert1}) and (\ref{pert2}) in order to decompose the scalar
type perturbations into curvature and entropy components and to get
their evolution equations. In subsequent sections, applied to
inflation, the evolution dynamics of perturbations will be studied
under the slow-roll approximation.

For our following study, it is essential to notice that equation
(\ref{pert1}) can be recast in the form
\begin{equation}\label{dzeta}
\dot{\mathcal{R}}=\left(\ln\frac{\dot{\varphi}^2}{2F\dot{\varphi}^2+3\dot{F}^2}\right)^{\bullet}\frac{2HF+\dot{F}}{2F\dot{\varphi}^2+3\dot{F}^2}\delta s+\frac{2HF+\dot{F}}{2F\dot{\varphi}^2+3\dot{F}^2}\frac{F}{a^2}\nabla^2(\phi+\psi),
\end{equation}
with the curvature perturbation
\begin{equation}\label{zeta}
\mathcal{R}=\frac{1}{2}(\phi+\psi)+\frac{2HF+\dot{F}}{2F\dot{\varphi}^2+3\dot{F}^2}\left[F(\dot{\phi}+\dot{\psi})+\frac{1}{2}(2HF+\dot{F})(\phi+\psi)\right].
\end{equation}
The first term on the right hand side of (\ref{dzeta}) can be
taken as the entropy perturbation, \footnote{Here is an ambiguity
in normalizing entropy perturbation. We define the entropy
perturbation by (\ref{entropy}), but we will abuse it for $\delta
s$ and $\delta\tilde{s}$ since they are proportional to
$\mathcal{S}$ up to background quantities. Through relations
(\ref{entropy}) and (\ref{std}) they are easy to be traded with
each other. Our normalization of (\ref{entropy}) is chosen to
ensure that
$\mathcal{P}_{\mathcal{R}\ast}=\mathcal{P}_{\mathcal{S}\ast}$ when
perturbations cross the Hubble horizon, as given by equation
(\ref{spectra-ast}).} or more exactly, the relative entropy
perturbation \cite{Gordon:2000hv}
\begin{equation}\label{entropy}
\mathcal{S}=\frac{\dot{F}(2HF+\dot{F})}{\dot{\varphi}(2F\dot{\varphi}^2+3\dot{F}^2)}\sqrt{\frac{3}{2F}}\delta s,
\end{equation}
where
\begin{eqnarray}\label{entrpert}
\delta s&=&F(\dot{\phi}+\dot{\psi})+\frac{1}{2}(2HF+\dot{F})(\phi+\psi)+\frac{2F\dot{\varphi}^2+3\dot{F}^2}{2\dot{F}}(\phi-\psi).
\end{eqnarray}
Strictly speaking, the second term on the right hand side of
(\ref{dzeta}) also contributes to the total entropy perturbation,
but it is suppressed on super-horizon scale \cite{Gordon:2000hv}.
Throughout our paper, we focus on the relative entropy
perturbation, and refer it as entropy perturbation for simplicity.
The adiabatic (curvature) and isocurvature (entropy) perturbations
were investigated in
\cite{GarciaBellido:1995qq,GrootNibbelink:2001qt,DiMarco:2002eb,DiMarco:2005nq}
for a two-field Lagrangian with a non-standard kinetic term. The
Lagrangian discussed there is equivalent to the $f(\varphi,R)$
theory here, up to a conformal transformation
\cite{Maeda:1988ab,Hwang:1996np}. One can check that the curvature
perturbation defined in (\ref{zeta}) is conformally equivalent to
that appeared in
\cite{GarciaBellido:1995qq,GrootNibbelink:2001qt,DiMarco:2002eb,DiMarco:2005nq}.
See also appendix \ref{app-conf}. As another check, in the
minimally coupled limit $f=M_p^2R$, the expression (\ref{zeta})
reduces to the familiar form
$\mathcal{R}=\phi-H(\dot{\phi}+H\phi)/\dot{H}$.

Along the line of \cite{Bardeen:1980kt}, we give an apparently
more traditional definition of curvature perturbation
$\mathcal{R}_{eff}$ and entropy perturbation $\delta s_{eff}$ in
appendix \ref{app-perts}. At the first glance, the traditional
definition seems more physical. But it depends heavily on an
artificial separation of gravity and matter content. The choice
given by (\ref{zeta}) and (\ref{entrpert}) is more convenient in
calculation. Moreover, the curvature and entropy perturbations at
the end of inflation are not necessarily the ones probed by
astronomical observations \cite{Komatsu:2008hk,:2006uk}, because
they may evolve significantly after the exit of inflation,
depending on the details of reheating. We will take a pragmatic
attitude and prefer the convenient definition (\ref{zeta}) and
(\ref{entrpert}) here.

One may still raise a question: why do we claim that $\mathcal{R}$
and $\mathcal{S}$ defined above correspond to curvature perturbation
and entropy perturbation respectively? This doubt can be resolved by
rewriting (\ref{zeta}) and (\ref{entrpert}) into the following form:
\begin{eqnarray}\label{zeta-ds}
\nonumber \mathcal{R}&=&\psi+\frac{2HF+\dot{F}}{2F\dot{\varphi}^2+3\dot{F}^2}\dot{\varphi}\delta\varphi+\frac{3H\dot{F}-\dot{\varphi}^2}{2F\dot{\varphi}^2+3\dot{F}^2}\delta F,\\
\delta s&=&\dot{\varphi}^2\left(\frac{\delta\varphi}{\dot{\varphi}}-\frac{\delta F}{\dot{F}}\right).
\end{eqnarray}
When deriving these relations, we have used equations
(\ref{Friedmann2}), (\ref{g0i}) and (\ref{gij}).

First of all, with (\ref{zeta-ds}) at hand, we can apply it to the
special limits mentioned in section \ref{sect-intro}:
\begin{itemize}
\item $f(R)$ models without a scalar field;
\item $F(\varphi)R$ scalar-tensor theory.
\end{itemize}
These models are extensively studied in the past
\cite{Hwang:1990re,Hwang:1996np,Hwang:1996bc,Hwang:1997uc,Hwang:2005hb}.
For $f(R)$ models, $\dot{\varphi}^2=0$, thus the entropy
perturbation vanishes obviously. As for scalar-tensor theory, since
$F$ becomes a function exclusively dependent of $\varphi$, we have
\begin{equation}
\frac{\delta F}{\dot{F}}=\frac{\delta\varphi}{\dot{\varphi}},
\end{equation}
which guarantees $\delta s=0$. Now it becomes very clear that there
is no entropy perturbation in these models. This is in agreement
with the fact that there is only one degree of freedom in these
models. On the other hand, curvature perturbation $\mathcal{R}$
reduces to
\begin{equation}
\mathcal{R}=\left\{
\begin{array}{ll}
\psi+\frac{H}{\dot{\varphi}}\delta\varphi,&\hbox{for pure $f(R)$ gravity without scalar field;}\\
\psi+\frac{H}{\dot{F}}\delta F,&\hbox{for $f=F(\varphi)R$.}
\end{array}
\right.
\end{equation}
These are exactly the ones having appeared in \cite{Hwang:2005hb}.
Our definition of curvature perturbation $\mathcal{R}$ naturally
generalizes them in a unified form.

Secondly, in favor of the small dictionary (\ref{gauge}), one can
prove that the definitions of $\mathcal{R}$ and $\delta s$ are
gauge-invariant. In other words, the $(\mathcal{B}-\mathcal{E}')$
terms cancel out automatically.

As a further check, since our definition of curvature perturbation
is gauge-invariant, by choosing a special (but not longitudinal)
gauge
\begin{equation}\label{comoving}
(2HF+\dot{F})\dot{\varphi}\delta\varphi^{(c)}+(3H\dot{F}-\dot{\varphi}^2)\delta F^{(c)}=0,
\end{equation}
we get a neat relation
\begin{equation}\label{curvpert}
\mathcal{R}=\psi^{(c)}.
\end{equation}
In minimally coupled models, the same relation (\ref{curvpert})
holds in comoving gauge or uniform density gauge
$\delta\varphi^{(c)}=0$. Here the gauge condition (\ref{comoving})
generalizes the comoving gauge condition, so we can take it as a
generalized comoving gauge. In fact, just as in minimal models,
relation (\ref{curvpert}) has a geometric interpretation. The
spatial curvature to the first order of perturbations is given by
\begin{equation}
^{(3)}R=\frac{4}{a^2}\psi^{(c)}.
\end{equation}
Therefore, in gauge (\ref{comoving}), the adiabatic perturbation
$\mathcal{R}$ is proportional to the spatial curvature $^{(3)}R$.
This is why we give it the name ``curvature perturbation''.

According to our definition in this section, the entropy
perturbation is proportional to
$\frac{\delta\varphi}{\dot{\varphi}}-\frac{\delta F}{\dot{F}}$. This
is natural if one remembers that a new scalar degree of freedom is
related to $F$ in the present theory. It gets dynamical when $F$
becomes dynamical. This form is very similar to the entropy
perturbation in models with two fields, say $\varphi_1$,
$\varphi_2$, corresponding to an entropy perturbation proportional
to
$\frac{\delta\varphi_1}{\dot{\varphi}_1}-\frac{\delta\varphi_2}{\dot{\varphi}_2}$.

We would like to emphasize that the perturbations defined in
appendix \ref{app-perts} make senses only if one views the
$f(\varphi,R)$ theory effectively as Einstein gravity with exotic
matter contents. This viewpoint is not so reliable since the gravity
has been modified in $f(\varphi,R)$ model. Furthermore, what we are
really interested in is the evolution of $\phi$ and $\psi$. As we
will show below, the choice in this section is powerful to study
their evolution.

In appendix \ref{app-entr}, we demonstrate that the evolution of
entropy perturbation obeys the equation
\begin{eqnarray}\label{evol-s}
\nonumber \ddot{\delta s}&=&\left\{\left[\ln\frac{\dot{\varphi}^2(2F\dot{\varphi}^2+3\dot{F}^2)}{\dot{F}^2}\right]^{\bullet}-3H\right\}\dot{\delta s}+\frac{\nabla^2}{a^2}\delta s\\
\nonumber &&+\Biggl\{\left(\frac{F\dot{\varphi}^2}{\dot{F}}\right)^{\bullet\bullet}+\left(\frac{3\dot{F}\ddot{\varphi}}{\dot{\varphi}}-\dot{\varphi}^2-\frac{3}{2}\ddot{F}\right)^{\bullet}+\left(\frac{2\ddot{\varphi}}{\dot{\varphi}}-\frac{3\dot{F}}{2F}\right)\left(\frac{3\dot{F}\ddot{\varphi}}{\dot{\varphi}}-\dot{\varphi}^2-3\ddot{F}\right)\\
\nonumber &&+\left(\frac{\dot{\varphi}^2}{\dot{F}}+\frac{3\dot{F}}{2F}\right)\left[2F(2H^2+\dot{H})+\frac{3\dot{F}\ddot{\varphi}}{\dot{\varphi}}-\dot{\varphi}^2-3\ddot{F}-\frac{F^2}{3F_{,R}}-\frac{F\dot{F}F_{,\varphi}}{2F_{,R}\dot{\varphi}}\right]\\
\nonumber &&+\left[\left(\ln\frac{\dot{\varphi}^2(2F\dot{\varphi}^2+3\dot{F}^2)}{\dot{F}^2}\right)^{\bullet}-3H\right]\left[\dot{\varphi}^2+\frac{3}{2}\ddot{F}-\left(\frac{F\dot{\varphi}^2}{\dot{F}}\right)^{\bullet}-\frac{3\dot{F}\ddot{\varphi}}{\dot{\varphi}}\right]\Biggr\}\\
\nonumber &&\times\frac{2\dot{F}}{2F\dot{\varphi}^2+3\dot{F}^2}\delta s\\
&&+\left[\frac{2F\dot{\varphi}^2}{3\dot{F}}+F\left(\ln\frac{\dot{F}^2}{2F\dot{\varphi}^2+3\dot{F}^2}\right)^{\bullet}\right]\frac{\nabla^2}{a^2}(\phi+\psi).
\end{eqnarray}
In terms of
\begin{equation}\label{std}
\delta\tilde{s}=\frac{\dot{F}}{\dot{\varphi}\sqrt{4F\dot{\varphi}^2+6\dot{F}^2}}\delta s,
\end{equation}
this equation can be reexpressed as
\begin{eqnarray}\label{evol-std}
\nonumber &&\ddot{\delta\tilde{s}}+3H\dot{\delta\tilde{s}}-\frac{\nabla^2}{a^2}\delta\tilde{s}\\
\nonumber &=&\frac{1}{2}\left[\ln\frac{\dot{\varphi}^2(2F\dot{\varphi}^2+3\dot{F}^2)}{\dot{F}^2}\right]^{\bullet}\Biggl\{\left[\frac{3}{2}\ln\frac{\dot{\varphi}^2(2F\dot{\varphi}^2+3\dot{F}^2)}{\dot{F}^2}\right]^{\bullet}\\
\nonumber &&-3H-\left[\ln\left(\frac{\dot{\varphi}^2(2F\dot{\varphi}^2+3\dot{F}^2)}{\dot{F}^2}\right)^{\bullet}\right]^{\bullet}\Biggr\}\delta\tilde{s}\\
\nonumber &&+\Biggl\{\left(\frac{F\dot{\varphi}^2}{\dot{F}}\right)^{\bullet\bullet}+\left(\frac{3\dot{F}\ddot{\varphi}}{\dot{\varphi}}-\dot{\varphi}^2-\frac{3}{2}\ddot{F}\right)^{\bullet}+\left(\frac{2\ddot{\varphi}}{\dot{\varphi}}-\frac{3\dot{F}}{2F}\right)\left(\frac{3\dot{F}\ddot{\varphi}}{\dot{\varphi}}-\dot{\varphi}^2-3\ddot{F}\right)\\
\nonumber &&+\left(\frac{\dot{\varphi}^2}{\dot{F}}+\frac{3\dot{F}}{2F}\right)\left[2F(2H^2+\dot{H})+\frac{3\dot{F}\ddot{\varphi}}{\dot{\varphi}}-\dot{\varphi}^2-3\ddot{F}-\frac{F^2}{3F_{,R}}-\frac{F\dot{F}F_{,\varphi}}{2F_{,R}\dot{\varphi}}\right]\\
\nonumber &&+\left[\left(\ln\frac{\dot{\varphi}^2(2F\dot{\varphi}^2+3\dot{F}^2)}{\dot{F}^2}\right)^{\bullet}-3H\right]\left[\dot{\varphi}^2+\frac{3}{2}\ddot{F}-\left(\frac{F\dot{\varphi}^2}{\dot{F}}\right)^{\bullet}-\frac{3\dot{F}\ddot{\varphi}}{\dot{\varphi}}\right]\Biggr\}\\
\nonumber &&\times\frac{2\dot{F}}{2F\dot{\varphi}^2+3\dot{F}^2}\delta\tilde{s}\\
&&+\left[\frac{2F\dot{\varphi}^2}{3\dot{F}}+F\left(\ln\frac{\dot{F}^2}{2F\dot{\varphi}^2+3\dot{F}^2}\right)^{\bullet}\right]\frac{\dot{F}}{\dot{\varphi}\sqrt{4F\dot{\varphi}^2+6\dot{F}^2}}\frac{\nabla^2}{a^2}(\phi+\psi).
\end{eqnarray}
At the same time, according to (\ref{dzeta}), the evolution equation
of $\mathcal{R}$ is
\begin{equation}\label{evol-zita}
\dot{\mathcal{R}}=\left(\ln\frac{\dot{\varphi}^2}{2F\dot{\varphi}^2+3\dot{F}^2}\right)^{\bullet}\frac{2\dot{\varphi}(2HF+\dot{F})}{\dot{F}\sqrt{4F\dot{\varphi}^2+6\dot{F}^2}}\delta\tilde{s}+\frac{2HF+\dot{F}}{2F\dot{\varphi}^2+3\dot{F}^2}\frac{F}{a^2}\nabla^2(\phi+\psi),
\end{equation}

There are similar equations in non-standard two-field models
\cite{GarciaBellido:1995qq,GrootNibbelink:2001qt,DiMarco:2002eb,DiMarco:2005nq}.
In principle, equations (\ref{evol-std}) and (\ref{evol-zita}) may
be obtained by a conformal transformation from the counterparts in
\cite{GarciaBellido:1995qq,GrootNibbelink:2001qt,DiMarco:2002eb,DiMarco:2005nq}.
However, having avoided the intricacies of conformal
transformation at the perturbation level, our derivation in the
$f(\varphi,R)$ frame is straightforward. Nonetheless, it is still
interesting to compare the results here and those in
\cite{GarciaBellido:1995qq,GrootNibbelink:2001qt,DiMarco:2002eb,DiMarco:2005nq}
via the conformal transformation \cite{Maeda:1988ab,Hwang:1996np}.
Such a comparison would confirm the conformal equivalence at the
perturbation level.

Until now we have not made any approximation. Equations
(\ref{evol-std}) and (\ref{evol-zita}) determine the dynamics of
entropy and curvature perturbations exactly. They are applicable in
various cosmological stages and scenarios of FLRW universe. Given a
concrete model, the classical evolution of perturbations can be
numerically followed utilizing these equations. To work in the
inflation scenario and generate an appropriate spectrum of density
fluctuation, we should consider the generalized slow-roll conditions
and the initial condition. This is a task of the next section.

\section{Slow-roll Approximation and Quantization}\label{sect-slroll}
In this section we will study the perturbations under the
generalized slow-roll approximation. Firstly we study the classical
evolution. At the end of this section, we will discuss the
quantization of perturbations as an initial condition.

From equation (\ref{zeta}), we know that the curvature perturbation
$\mathcal{R}$ is fully determined by $\phi+\psi$ and its time
derivative. So the dynamics of $\phi+\psi$ informs us the dynamics
of $\mathcal{R}$. Inserting (\ref{zeta}) into (\ref{dzeta}), one
obtains
\begin{eqnarray}\label{dels-plus}
\nonumber \frac{1}{F}\left(\ln\frac{2F\dot{\varphi}^2+3\dot{F}^2}{\dot{\varphi}^2}\right)^{\bullet}\delta s+(\ddot{\phi}+\ddot{\psi})+\left(\ln\frac{aF^3}{2F\dot{\varphi}^2+3\dot{F}^2}\right)^{\bullet}(\dot{\phi}+\dot{\psi})&&\\
+\frac{1}{2}[\ln(a^2F)]^{\bullet}\left[\ln\frac{(2HF+\dot{F})^2}{2F\dot{\varphi}^2+3\dot{F}^2}\right]^{\bullet}(\phi+\psi)-\frac{\nabla^2}{a^2}(\phi+\psi)&=&0.
\end{eqnarray}
We observe that substituting this equation into (\ref{evol-s}) will
result in a fourth order differential equation of $\phi+\psi$. But
that equation would be rather difficult to solve directly. In this
section, we rewrite equations (\ref{evol-s}) and (\ref{dels-plus})
into a tractable form. In the next section, we will get their
analytical solutions in some special examples.

Firstly, let us define two new variables:
\begin{equation}
u_{\mathcal{R}}=\frac{F^\frac{3}{2}}{\sqrt{4F\dot{\varphi}^2+6\dot{F}^2}}(\phi+\psi),~~~~u_{\mathcal{S}}=a\delta\tilde{s}.
\end{equation}
Similar to $\phi+\psi$, the variable $u_{\mathcal{R}}$ tells us the
evolution of curvature perturbation $\mathcal{R}$. Following
relation (\ref{dels-plus}), it obeys the equation of motion
\begin{equation}
\frac{a\dot{\varphi}F^{\frac{1}{2}}}{\dot{F}}\left(\ln\frac{2F\dot{\varphi}^2+3\dot{F}^2}{\dot{\varphi}^2}\right)^{\bullet}u_{\mathcal{S}}+u_{\mathcal{R}}''-\nabla^2u_{\mathcal{R}}+m_{\mathcal{R}}^2a^2u_{\mathcal{R}}=0,
\end{equation}
with the ``mass squared''
\begin{eqnarray}
\nonumber m_{\mathcal{R}}^2&=&\frac{F^\frac{3}{2}}{\sqrt{2F\dot{\varphi}^2+3\dot{F}^2}}\left(\frac{\sqrt{2F\dot{\varphi}^2+3\dot{F}^2}}{F^\frac{3}{2}}\right)^{\bullet\bullet}-\frac{1}{2}\left(\ln\frac{2F\dot{\varphi}^2+3\dot{F}^2}{aF^3}\right)^{\bullet}\left(\ln\frac{2F\dot{\varphi}^2+3\dot{F}^2}{F^3}\right)^{\bullet}\\
&&+\frac{1}{2}[\ln(a^2F)]^{\bullet}\left[\ln\frac{(2HF+\dot{F})^2}{2F\dot{\varphi}^2+3\dot{F}^2}\right]^{\bullet}.
\end{eqnarray}
If we introduce the notation
\begin{equation}
E=\frac{2HF+\dot{F}}{F^\frac{3}{2}},
\end{equation}
then one can easily prove that
\begin{equation}
\dot{E}=-\frac{2F\dot{\varphi}^2+3\dot{F}^2}{2F^\frac{5}{2}}.
\end{equation}
During the inflation stage, the Hubble parameter and the effective
energy density are almost constants. In the following, we will study
perturbation dynamics under the slow-roll approximation. For the
convenience of calculation, our definition of slow-roll parameters
is
\begin{eqnarray}\label{slroll-para}
\nonumber &&\epsilon_1=\frac{\dot{H}}{H^2},~~~~\epsilon_2=\frac{\ddot{H}}{H\dot{H}},~~~~\eta_1=\frac{\ddot{\varphi}}{H\dot{\varphi}},\\
\nonumber &&\eta_2=\frac{\dddot{\varphi}}{H\ddot{\varphi}},~~~~\delta_1=\frac{\dot{F}}{HF},~~~~\delta_2=\frac{\dot{E}}{HE},\\
&&\delta_3=\frac{\ddot{F}}{H\dot{F}},~~~~\delta_4=\frac{\ddot{E}}{H\dot{E}},~~~~\delta_6=\frac{\dddot{E}}{H\ddot{E}}~.
\end{eqnarray}
While the slow-roll conditions are given by
\begin{equation}\label{slroll-cond}
|\epsilon_i|\ll1,~~~~|\eta_i|\ll1,~~~~|\delta_i|\ll1.
\end{equation}
Be careful with the notation and sign difference between the
slow-roll parameters here and those in most literature. Under the
slow-roll conditions, the background equations
(\ref{Friedmann1}-\ref{eom-varphi}) are significantly simplified,
\begin{eqnarray}
\label{simp-Friedmann1}&&V-\frac{1}{2}f+3H^2F\simeq0,\\
\label{simp-Friedmann2}&&\dot{\varphi}^2+2\dot{H}F-H\dot{F}\simeq0,\\
\label{simp-eom-varphi}&&3H\dot{\varphi}-\frac{1}{2}f_{,\varphi}+V_{,\varphi}\simeq0.
\end{eqnarray}
Notice that equation (\ref{simp-eom-varphi}) can be consistently
derived from the leading order Friedmann equations
(\ref{simp-Friedmann1}) and (\ref{simp-Friedmann2}). Neglecting
higher order terms, we have
\begin{eqnarray}
\nonumber &&\dot{E}=-\frac{\dot{\varphi}^2}{F^\frac{3}{2}},~~~~\ddot{E}=\frac{3\dot{F}\dot{\varphi}^2-4F\dot{\varphi}\ddot{\varphi}}{2F^\frac{5}{2}},~~~~\delta_2\simeq\epsilon_1-\frac{1}{2}\delta_1,\\
&&\delta_4\simeq2\eta_1-\frac{3}{2}\delta_1,~~~~\delta_6\simeq\eta_1-\frac{5}{2}\delta_1+\frac{3\delta_1\delta_3-\delta_1\eta_1-4\eta_1\eta_2}{3\delta_1-4\eta_1}.
\end{eqnarray}

Under the slow-roll approximation, the Hubble parameter is almost
constant during inflation. In terms of slow-roll parameters, the
nondimensionalized ``mass squared'' for $u_{\mathcal{R}}$ is
\begin{eqnarray}
\frac{m_{\mathcal{R}}^2}{H^2}&\simeq&2\epsilon_1-\eta_1.
\end{eqnarray}
Therefore, to the leading order, the evolution equation of
$u_{\mathcal{R}}$ is
\begin{equation}\label{evol-uzita}
u''_{\mathcal{R} k}+k^2u_{\mathcal{R} k}+m_{\mathcal{R}}^2a^2u_{\mathcal{R} k}+\beta u_{\mathcal{S}k}=0
\end{equation}
in the Fourier space. Here we have used the notation
\begin{eqnarray}
\beta&\simeq&\sign(\dot{\varphi})aH\sqrt{\delta_1-2\epsilon_1}
\end{eqnarray}
with $\sign(\dot{\varphi})=\dot{\varphi}/|\dot{\varphi}|$. In other
words, the sign of $\beta$ depends on the sign of $\dot{\varphi}$.

At the same time, in terms of $u_{\mathcal{S}}$, the evolution equation
(\ref{evol-std}) of entropy perturbation can be written to the
leading order as
\begin{equation}\label{evol-uds}
u''_{\mathcal{S}k}+k^2u_{\mathcal{S}k}+m_{\mathcal{S}}^2a^2u_{\mathcal{S}k}+\alpha k^2u_{\mathcal{R} k}=0,
\end{equation}
in which
\begin{eqnarray}
\label{mass-ds}\frac{m_{\mathcal{S}}^2}{H^2}&\simeq&\frac{5}{2}\delta_1-5\epsilon_1+\frac{F}{3H^2F_{,R}}+\frac{\dot{F}F_{,\varphi}}{2H^2F_{,R}\dot{\varphi}}-6,\\
\frac{\alpha}{aH}&=&\frac{\dot{F}}{H\dot{\varphi}F^{\frac{1}{2}}}\left[\frac{2\dot{\varphi}^2}{3\dot{F}}+\left(\ln\frac{\dot{F}^2}{2F\dot{\varphi}^2+3\dot{F}^2}\right)^{\bullet}\right]\simeq\sign(\dot{\varphi})\frac{2}{3}\sqrt{\delta_1-2\epsilon_1}.
\end{eqnarray}

In the above, we discussed the classical perturbations and a
generalized slow-roll condition. The initial condition is governed
by quantum theory of fluctuations. That is, we should expand action
(\ref{action}) to the second order with respect to perturbations,
taking the form
\begin{equation}\label{quantization}
\delta_2S=\int d^3\vec{x}d\tau\left[\frac{1}{2}(\partial_\tau v_{\mathcal{R}})^2+\frac{1}{2}(\partial_\tau v_{\mathcal{S}})^2-\frac{1}{2}(\partial_iv_{\mathcal{R}})^2-\frac{1}{2}(\partial_iv_{\mathcal{S}})^2+C_Jv_{\mathcal{R}}\partial_\tau v_{\mathcal{S}}-V(v_{\mathcal{R}},v_{\mathcal{S}})\right].
\end{equation}
In this action the covariant variables $v_{\mathcal{R}}$ and $v_{\mathcal{S}}$ are
linear superpositions of perturbations. Then the canonical
quantization leads to the initial condition
\begin{equation}\label{initialv}
v_{\mathcal{R} k}\rightarrow\frac{1}{\sqrt{2k}}e^{-ik\tau}\hat{e}_{\mathcal{R} k},~~~~v_{\mathcal{S}k}\rightarrow\frac{1}{\sqrt{2k}}e^{-ik\tau}\hat{e}_{\mathcal{S}k}
\end{equation}
in the short wavelength limit $k/(aH)\rightarrow\infty$. Here
$\{\hat{e}_{\mathcal{R} k},\hat{e}_{\mathcal{S}k}\}$ is the orthogonal
basis
\begin{equation}
\langle\hat{e}_{\alpha k},\hat{e}_{\beta k'}\rangle=\delta_{\alpha\beta}\delta(k-k'),~~~~\alpha,\beta=\mathcal{R},\mathcal{S}.
\end{equation}
From the argument in appendix \ref{app-conf}, we believe the
canonical variables in $f(\varphi,R)$ gravity should be
\begin{eqnarray}\label{quant-var}
\nonumber v_{\mathcal{R}}&=&a\sqrt{\frac{2F}{2F\dot{\varphi}^2+3\dot{F}^2}}\left[\dot{\varphi}\delta\varphi+3\dot{F}\psi+\frac{F(\dot{\varphi}^2-3H\dot{F})}{2HF+\dot{F}}(\phi+\psi)\right]\\
\nonumber &=&a\sqrt{\frac{2F}{2F\dot{\varphi}^2+3\dot{F}^2}}\frac{2F\dot{\varphi}^2+3\dot{F}^2}{2HF+\dot{F}}\mathcal{R},\\
v_{\mathcal{S}}&=&\frac{\sqrt{3}a}{\sqrt{2F\dot{\varphi}^2+3\dot{F}^2}}[\dot{F}\delta\varphi+F\dot{\varphi}(\phi-\psi)]=\frac{\sqrt{3}a\dot{F}}{\dot{\varphi}\sqrt{2F\dot{\varphi}^2+3\dot{F}^2}}\delta s=\sqrt{6}u_{\mathcal{S}}.
\end{eqnarray}
A detailed form of (\ref{quantization}) and a closer analysis of
canonical quantization will be presented elsewhere.

When solving equations (\ref{evol-uzita}) and (\ref{evol-uds}), the
initial condition (\ref{initialv}) for $v_{\mathcal{R}}$ and
$v_{\mathcal{S}}$ sets the initial condition of $u_{\mathcal{R}}$ and
$u_{\mathcal{S}}$. In the short wavelength limit, we find
$v_{\mathcal{R} k}\rightarrow2u'_{\mathcal{R} k}$. Thus we have
\begin{equation}\label{initialu}
u_{\mathcal{R} k}\rightarrow\frac{1}{(2k)^{\frac{3}{2}}}e^{i\left(\frac{\pi}{2}-k\tau\right)}\hat{e}_{\mathcal{R} k},~~~~u_{\mathcal{S}k}\rightarrow\frac{1}{\sqrt{12k}}e^{-ik\tau}\hat{e}_{\mathcal{S}k}.
\end{equation}
One can check that this solution meets (\ref{evol-uzita}) and
(\ref{evol-uds}) well in the limit $k/(aH)\rightarrow\infty$.

\section{Analytical Examples}\label{sect-example}
In equations (\ref{evol-uzita}) and (\ref{evol-uds}), it is clear
that $u_{\mathcal{R} k}$ and $u_{\mathcal{S}k}$ are coupled, while the
coupling strength is controlled by $\alpha$ and $\beta$. In general
the coupled equations are difficult to solve analytically, thus
numerical approaches are needed. Nevertheless, under some
circumstances, analytical results are available. Let us focus on the
case $\delta s=0$ in subsection \ref{subsect-noen}. In subsection
\ref{subsect-gR2}, we discuss the $g(\varphi)R^2$ model with a
non-vanishing entropy perturbation. Under certain approximation, we
solve a special case analytically, resulting in nearly
scale-invariant power spectra.

\subsection{No Relative Entropy Perturbation: $\delta s=0$}\label{subsect-noen}
This is essentially the case with only one degree of freedom.
Therefore, when $\delta s=0$, equation (\ref{evol-uds}) is
expected to be satisfied automatically, while the dynamics of the
survival degree of freedom is described by equation
(\ref{evol-uzita}). Although we cannot give a general proof to
this claim, we can check it in two classes of models mentioned
previously:
\begin{itemize}
\item $f(R)$ models without a scalar field;
\item $F(\varphi)R$ scalar-tensor theory.
\end{itemize}
For $f(R)$ models, this is quite clear because  $u_{\mathcal{S}}=0$
and $\alpha=0$. For $F(\varphi)R$ scalar-tensor theory, it is less
obvious. But looking at equation (\ref{mass-ds}), one may notice
that we have assumed $F_{,R}\neq0$ when writing down
(\ref{evol-uds}) and there is a singularity in $m_{\mathcal{S}}^2$.
Multiplying equation (\ref{evol-uds}) with $F_{,R}$ to eliminate the
illegal singularity in $m_{\mathcal{S}}^2$, we will find the
resulting equation is well-defined and automatically satisfied.

We can not exclude the possibility that the above claim does not
hold in some cases, although so far we can not come up with such
an example. In those cases, equation (\ref{evol-uds}) gives
$u_{\mathcal{R}}=0$, which also satisfies (\ref{evol-uzita}). But
then there will be no quantum fluctuations. So we are not
interested in this possibility even if it exists.

Hence let us assume equation (\ref{evol-uds}) is satisfied more
generally in the cases we are interested with $\delta s=0$. Then
we are left with one differential equation of $u_{\mathcal{R}}$,
\begin{equation}\label{ex1-evol-uzita}
u''_{\mathcal{R} k}+k^2u_{\mathcal{R} k}+m_{\mathcal{R}}^2a^2u_{\mathcal{R} k}=0,
\end{equation}
which is obtained by setting  $u_{\mathcal{S}}=0$ in equation
(\ref{evol-uzita}).

In accordance with initial condition (\ref{initialu}), we find the
solution for (\ref{ex1-evol-uzita}) is
\begin{equation}\label{ex1-sol-uzitak}
u_{\mathcal{R} k}=-\frac{1}{4k^{\frac{3}{2}}}e^{i(\nu-\frac{1}{2})\frac{\pi}{2}}\sqrt{-\pi k\tau}H^{(1)}_{\nu}(-k\tau)=C\sqrt{z}H^{(1)}_{\nu}(z),~~~~\mathrm{with}~\nu^2=\frac{1}{4}-\frac{m_{\mathcal{R}}^2}{H^2}.
\end{equation}
Here we introduced the notation $z=-k\tau$. This form of solution is
consistent with the minimal inflation model \cite{Riotto:2002yw}.
For $\nu\simeq\frac{1}{2}$, it gives a nearly scale-invariant power
spectrum. The coefficient $C$ is determined by quantization
condition (\ref{initialu}). Keep in mind that in the limit
$z\rightarrow\infty$,
\begin{equation}
H_{\nu}^{(1)}(z)\rightarrow\sqrt{\frac{2}{\pi z}}e^{i(z-\frac{\nu\pi}{2}-\frac{\pi}{4})}\propto e^{-ik\tau}.
\end{equation}
According to inflation theory, the CMB fluctuations are seeded by
primordial quantum fluctuations inside the horizon in the early
universe. When the primordial fluctuations crossed the Hubble
horizon, they began the classical evolution phase. So the initial
condition for classical evolution is dictated by quantization. Here
we did not study the quantization condition carefully. It would be
important to put this result on firmer ground by quantizing the
second order action directly.

If the scale factor is exponentially growing $a\sim e^{Ht}$, then
there is a relation
\begin{equation}
aH=-\frac{1}{\tau}=\frac{k}{z}.
\end{equation}
Straightforward calculations give
\begin{eqnarray}
\nonumber u'_{\mathcal{R} k}&=&-Ck\left[\sqrt{z}H^{(1)}_{\nu-1}(z)+\frac{1-2\nu}{2\sqrt{z}}H^{(1)}_{\nu}(z)\right],\\
u''_{\mathcal{R} k}&=&-Ck^2\left(\sqrt{z}+\frac{1-4\nu^2}{4z^{\frac{3}{2}}}\right)H^{(1)}_{\nu}(z).
\end{eqnarray}
Notice that a prime denotes the derivative with respect to conformal
time $\tau$.

In the long wavelength limit $k\tau\rightarrow0$, for
$\nu\simeq\frac{1}{2}$, we obtain to the leading order of slow-roll
parameters,
\begin{eqnarray}
\nonumber \mathcal{R}_k&\simeq&\frac{1}{4k^\frac{3}{2}}\left(\frac{2HF+\dot{F}}{F}\right)^2\sqrt{\frac{F}{2F\dot{\varphi}^2+3\dot{F}^2}}\left(\frac{-k\tau}{2}\right)^{\frac{1}{2}-\nu}\\
&\simeq&\frac{H^2}{k^\frac{3}{2}}\sqrt{\frac{F}{2F\dot{\varphi}^2+3\dot{F}^2}}.
\end{eqnarray}
As a result, the power spectrum of curvature perturbation is
\begin{equation}
\mathcal{P}_{\mathcal{R}}=\frac{k^3}{2\pi^2}|\mathcal{R}_k|^2=\frac{H^4F}{2\pi^2(2F\dot{\varphi}^2+3\dot{F}^2)},
\end{equation}
and thus the spectral index is
\begin{equation}\label{spectrumw}
n_{\mathcal{R}}-1=\frac{4\dot{H}}{H^2}+\frac{\dot{F}}{HF}-\frac{(2F\dot{\varphi}^2+3\dot{F}^2)^{\bullet}}{H(2F\dot{\varphi}^2+3\dot{F}^2)}~.
\end{equation}
Recalling that in \cite{Hwang:2005hb} Hwang and Noh gave the
corresponding result
\begin{equation}\label{spectrumh}
n_{\mathcal{R}}-1\simeq\left\{
\begin{array}{ll}
\frac{4\dot{H}}{H^2}+\frac{\dot{F}}{HF}-\frac{2\ddot{F}}{H\dot{F}},&\hbox{for pure $f(R)$ gravity without scalar field;} \\
\frac{4\dot{H}}{H^2}+\frac{\dot{F}}{HF}-\frac{(2F\dot{\varphi}^2+3\dot{F}^2)^{\bullet}}{H(2F\dot{\varphi}^2+3\dot{F}^2)},&\hbox{for $f=F(\varphi)R$.}
\end{array}
\right.
\end{equation}
Obviously the final results (\ref{spectrumw}) and (\ref{spectrumh})
are perfectly matched. Notice that for the case of pure $f(R)$
gravity, we can recover (\ref{spectrumh}) from (\ref{spectrumw}) by
setting $\dot{\varphi}=0$.

In the leading order, we also find $n_{\mathcal{R}}-1=1-2\nu$.
This indicates $\dot{\mathcal{R}}_k/(H\mathcal{R}_k)\simeq0$ for
long wavelength perturbations. That is to say, under the
generalized slow-roll approximation, the curvature perturbation is
almost conserved outside the horizon. On the one hand, this
extends the previous result in \cite{Hwang:2005hb} to the general
case $\delta s=0$ (and equation (\ref{evol-uds}) satisfied
automatically). On the other hand, to get richer phenomena, we
should take the entropy perturbation into account and solve
evolution equations (\ref{evol-uzita}) and (\ref{evol-uds}) more
generally.

\subsection{$g(\varphi)R^2$ Correction}\label{subsect-gR2}
A relatively simple but non-trivial model with non-vanishing entropy
perturbation is to consider the $R^2$ correction with a
$\varphi$-dependent coefficient,
\begin{equation}
f(\varphi,R)=M_p^2R+g(\varphi)R^2.
\end{equation}

For this class of model, under the slow-roll approximation,
equations (\ref{simp-Friedmann1}-\ref{simp-eom-varphi}) become
\begin{equation}
V\simeq3M_p^2H^2,~~~~\dot{\varphi}^2+2\dot{H}F-H\dot{F}\simeq0,~~~~3H\dot{\varphi}\simeq72g_{,\varphi}H^4-V_{,\varphi}.
\end{equation}
Then we get the following relations:
\begin{eqnarray}
\nonumber &&F=M_p^2+2gR\simeq M_p^2+\frac{8gV}{M_p^2},\\
\nonumber &&V_{,\varphi}\dot{\varphi}\simeq6M_p^2H\dot{H}=2HV\epsilon_1,\\
\nonumber &&V_{,\varphi}\ddot{\varphi}+V_{,\varphi\varphi}\dot{\varphi}^2\simeq6M_p^2(H\ddot{H}+\dot{H}^2),\\
\nonumber &&3H\ddot{\varphi}+3\dot{H}\dot{\varphi}\simeq\left(\frac{8g_{,\varphi}V^2}{M_p^4}-V_{,\varphi}\right)_{,\varphi}\dot{\varphi},\\
&&3H\dddot{\varphi}+6\dot{H}\ddot{\varphi}+3\ddot{H}\dot{\varphi}\simeq\left(\frac{8g_{,\varphi}V^2}{M_p^4}-V_{,\varphi}\right)_{,\varphi}\ddot{\varphi}+\left(\frac{8g_{,\varphi}V^2}{M_p^4}-V_{,\varphi}\right)_{,\varphi\varphi}\dot{\varphi}^2.
\end{eqnarray}
The slow-roll parameters (\ref{slroll-para}) can be expressed in
terms of $g$ and $V$ and their derivatives with respect to
$\varphi$,
\begin{eqnarray}
\nonumber &&\epsilon_1=\frac{\dot{H}}{H^2}\simeq\frac{4g_{,\varphi}V_{,\varphi}}{M_p^2}-\frac{M_p^2V_{,\varphi}^2}{2V^2},\\
\nonumber &&\eta_1=\frac{\ddot{\varphi}}{H\dot{\varphi}}\simeq\frac{8g_{,\varphi\varphi}V}{M_p^2}+\frac{12g_{,\varphi}V_{,\varphi}}{M_p^2}-\frac{M_p^2V_{,\varphi\varphi}}{V}+\frac{M_p^2V_{,\varphi}^2}{2V^2},\\
\nonumber &&\delta_1=\frac{\dot{F}}{HF}\simeq\frac{16\epsilon_1 V(gV)_{,\varphi}}{V_{,\varphi}(M_p^4+8gV)},~~~~\delta_2=\frac{\dot{E}}{HE}\simeq\epsilon_1-\frac{1}{2}\delta_1,\\
\nonumber &&\delta_3=\frac{\ddot{F}}{H\dot{F}}=\eta_1+\frac{2\epsilon_1 V(gV)_{,\varphi\varphi}}{V_{,\varphi}(gV)_{,\varphi}},~~~~\delta_4=\frac{\ddot{E}}{H\dot{E}}\simeq2\eta_1-\frac{3}{2}\delta_1,\\
\nonumber &&\epsilon_2=\frac{\ddot{H}}{H\dot{H}}\simeq\eta_1-\epsilon_1+\frac{2\epsilon_1 VV_{,\varphi\varphi}}{V_{,\varphi}^2},\\
\nonumber &&\eta_2=\frac{\dddot{\varphi}}{H\ddot{\varphi}}\simeq\eta_1-\epsilon_1-\frac{\epsilon_1\ddot{H}}{\eta_1 H\dot{H}}+\frac{2\epsilon_1 M_p^2}{\eta_1 V_{,\varphi}}\left(\frac{8g_{,\varphi}V^2}{M_p^4}-V_{,\varphi}\right)_{,\varphi\varphi},\\
&&\delta_6=\frac{\dddot{E}}{H\ddot{E}}\simeq\eta_1-\frac{5}{2}\delta_1+\frac{3\delta_1\delta_3-\delta_1\eta_1-4\eta_1\eta_2}{3\delta_1-4\eta_1}.
\end{eqnarray}

Subsequently, the coefficients in equations (\ref{evol-uzita}) and
(\ref{evol-uds}) are
\begin{eqnarray}
\nonumber &&m_{\mathcal{R}}^2\simeq H^2(2\epsilon_1-\eta_1),~~~~\beta\simeq\sign(\dot{\varphi})aH\sqrt{\delta_1-2\epsilon_1},~~~~\alpha\simeq\sign(\dot{\varphi})\frac{2}{3}aH\sqrt{\delta_1-2\epsilon_1},\\
&&m_{\mathcal{S}}^2\simeq H^2\left[\frac{5}{2}\delta_1-3\epsilon_1+\frac{M_p^4}{2gV}-2+\frac{48g_{,\varphi}(gV)_{,\varphi}}{gM_p^2}\right].
\end{eqnarray}

The above results are still too complicated to get some sense.
Particularly, since there are so many slow-roll parameters, one may
even worry about whether the slow-roll conditions
(\ref{slroll-cond}) can be satisfied simultaneously. However, for
the special case\footnote{As a matter of fact, a model with
$R^2\varphi^2$ correction was discussed in \cite{Li:2006te}.}
\begin{equation}\label{ex2-gV}
g(\varphi)=\frac{1}{4}\lambda\varphi^2,~~~~V(\varphi)=\frac{1}{2}m^2\varphi^2,
\end{equation}
the slow-roll parameters take much simpler form as below:
\begin{eqnarray}
\nonumber &&\epsilon_1=\frac{\dot{H}}{H^2}\simeq\frac{2\lambda m^2\varphi^2}{M_p^2}-\frac{2M_p^2}{\varphi^2},~~~~\eta_1=\frac{\ddot{\varphi}}{H\dot{\varphi}}\simeq\frac{8\lambda m^2\varphi^2}{M_p^2},\\
\nonumber &&\delta_1=\frac{\dot{F}}{HF}\simeq\frac{4\epsilon_1\lambda m^2\varphi^4}{M_p^4+\lambda m^2\varphi^4},~~~~\delta_2=\frac{\dot{E}}{HE}\simeq\epsilon_1-\frac{1}{2}\delta_1,\\
\nonumber &&\delta_3=\frac{\ddot{F}}{H\dot{F}}=\eta_1+3\epsilon_1,~~~~\delta_4=\frac{\ddot{E}}{H\dot{E}}\simeq2\eta_1-\frac{3}{2}\delta_1,\\
\nonumber &&\epsilon_2=\frac{\ddot{H}}{H\dot{H}}\simeq\eta_1,~~~~\eta_2=\frac{\dddot{\varphi}}{H\ddot{\varphi}}\simeq\eta_1+3\epsilon_1,\\
&&\delta_6=\frac{\dddot{E}}{H\ddot{E}}\simeq3\epsilon_1+2\eta_1-\frac{5}{2}\delta_1+\frac{\delta_1\eta_1}{4\eta_1-3\delta_1}.
\end{eqnarray}
We have been concentrating on the case $F>0$, so in this example we
limit our attention to the case with $\lambda>0$. Obviously the
slow-roll conditions (\ref{slroll-cond}) are satisfied when
$M_p^2\ll\varphi^2\ll M_p^2/(\lambda m^2)$. To meet this condition
we should fine-tune the coupling constants to be very small $\lambda
m^2\ll1$. This is the large field inflation. Although the value of
field $\varphi$ is larger than Planck mass, thanks to the small
coupling constants, its energy density is still less than the Planck
energy density.

In the above special case, if we further assume
$M_p^2/\varphi^2\ll\lambda m^2\varphi^2/M_p^2\ll1$ during inflation,
then we will find
\begin{eqnarray}\label{ex2-coe}
\nonumber &&\epsilon_1\simeq\frac{2\lambda m^2\varphi^2}{M_p^2},~~~~\eta_1\simeq\delta_1\simeq4\epsilon_1,\\
\nonumber &&m_{\mathcal{R}}^2\simeq-2\epsilon_1 H^2,~~~~\beta\simeq\sign(\dot{\varphi})aH\sqrt{2\epsilon_1},\\
&&m_{\mathcal{S}}^2\simeq(31\epsilon_1-2)H^2,~~~~\alpha\simeq\sign(\dot{\varphi})\frac{2}{3}aH\sqrt{2\epsilon_1}.
\end{eqnarray}
The model with coefficients (\ref{ex2-coe}) is relatively simple.
Let us discuss it in some detail.

Again, due to the non-vanishing $\alpha$ and $\beta$, the
interaction terms form an obstacle to our analytical study.
However, it is still interesting to make some rough estimates by
neglecting these interactions before horizon-crossing/Hubble-exit.
The problem is akin to the one we met in inflation of coupled
multiple field. In fact, there is an excellent analysis of
coupling effects in \cite{Byrnes:2006fr}. It turned out if the
coupling terms are of slow-roll order, then they will give a
correction of slow-roll order compared to the leading order
contribution. In our case, the coefficients of coupling terms are
of order $\mathcal{O}(\sqrt{\epsilon_1})$, so we expect their
corrections to the power spectra are suppressed by
$\mathcal{O}(\sqrt{\epsilon_1})$.

Our scheme is taking limit $M_p^2/\varphi^2\ll\lambda
m^2\varphi^2/M_p^2\ll1$, and disregarding the interaction terms
related to $\alpha$ and $\beta$ inside the horizon $k/(aH)\gtrsim1$.
In accordance with initial condition (\ref{initialu}), we get an
analytical solution to equations (\ref{evol-uzita}) and
(\ref{evol-uds}),
\begin{eqnarray}\label{ex2-sol-uzitak-udsk}
\nonumber &&u_{\mathcal{R} k}=-\frac{1}{4k^{\frac{3}{2}}}e^{i(\nu_1-\frac{1}{2})\frac{\pi}{2}}\sqrt{-\pi k\tau}H^{(1)}_{\nu_1}(-k\tau)\hat{e}_{\mathcal{R} k},~~~~\mathrm{with}~\nu_1^2=\frac{1}{4}-\frac{m_{\mathcal{R}}^2}{H^2},\\
&&u_{\mathcal{S}k}=-\frac{1}{2\sqrt{6k}}e^{i(\nu_2-\frac{3}{2})\frac{\pi}{2}}\sqrt{-\pi k\tau}H^{(1)}_{\nu_2}(-k\tau)\hat{e}_{\mathcal{S}k},~~~~\mathrm{with}~\nu_2^2=\frac{1}{4}-\frac{m_{\mathcal{S}}^2}{H^2}.
\end{eqnarray}
From this solution, we find at the time of horizon-crossing, the
power spectra of curvature and entropy perturbations are nearly
scale-invariant,
\begin{eqnarray}\label{spectra-ast}
\nonumber &&\mathcal{P}_{\mathcal{R}\ast}=\frac{k^3}{2\pi^2}|\mathcal{R}_{k\ast}|^2\simeq\left.\frac{H^4}{4\pi^2\dot{\varphi}^2}\right|_{\ast}\simeq\frac{M_p^2}{96\pi^2\lambda^2m^2\varphi^4_{\ast}},\\
&&\mathcal{P}_{\mathcal{S}\ast}=\frac{k^3}{2\pi^2}|\mathcal{S}_{k\ast}|^2\simeq\left.\frac{H^4}{4\pi^2\dot{\varphi}^2}\right|_{\ast}\simeq\frac{M_p^2}{96\pi^2\lambda^2m^2\varphi^4_{\ast}},
\end{eqnarray}
while their cross-correlation is negligible,\footnote{This is
because we have neglected the coupling terms in (\ref{evol-uzita})
and (\ref{evol-uds}) before Hubble-crossing. The fact is, when
taking coupling terms into consideration, we expect the
cross-correlation is not negligible here,
$\mathcal{P}_{\mathcal{C}\ast}/\mathcal{P}_{\mathcal{S}\ast}\sim\mathcal{O}(\sqrt{\epsilon_1})$.
But to get the explicit value of it, one should either perform the
higher order calculation or take a numerical method.}
\begin{equation}
\mathcal{P}_{\mathcal{C}\ast}=\frac{k^3}{2\pi^2}\langle\mathcal{R}_{k\ast},\mathcal{S}_{k\ast}\rangle\simeq0.
\end{equation}
Therefore the spectral indices at the horizon-crossing are
\begin{equation}
n_{\mathcal{R}\ast}-1=n_{\mathcal{S}\ast}-1=4\epsilon_{1\ast}-2\eta_{1\ast}=-4\epsilon_{1\ast}.
\end{equation}
The variables with an asterisk subscript take their values at the
horizon-crossing time $k=aH$. Note we have chosen the
normalization of $\mathcal{S}$ in (\ref{entropy}) so that
$\mathcal{P}_{\mathcal{R}\ast}=\mathcal{P}_{\mathcal{S}\ast}$.

We would like to pause here and comment on the scale invariance of
the entropy perturbation. If readers go through our calculation
carefully, they would find the limit $M_p^2/\varphi^2\ll\lambda
m^2\varphi^2/M_p^2\ll1$ is of key importance in making
$\mathcal{P}_{\mathcal{S}\ast}$ scale-invariant. In this limit,
the $M_p^4/(2gV)$ term is small, hence
$m_{\mathcal{S}}^2/H^2\simeq-2$ and subsequently $\nu_2\simeq3/2$.
If $M_p^4/(2gV)$ is not small enough, we cannot get such a value
of $\nu_2$ and then the entropy power spectrum will not be
scale-invariant. Even in the limit we have taken, there are
subtleties with the $M_p^4/(2gV)$ term. If
$M_p^4/(2gV)\sim\mathcal{O}(\epsilon_1)$, we should pick it up
when calculating $m_{\mathcal{S}}^2$. However, for simplicity of
our following estimation, when writing down (\ref{ex2-coe}), we
have assumed $M_p^4/(2gV)\ll\epsilon_1$.

If one considers interactions outside the horizon, it proves useful
to describe the evolution of perturbations in terms of a general
transfer matrix
\begin{equation}\label{trans-matrix}
\left(
\begin{array}{c}
\mathcal{R}\\
\mathcal{S}
\end{array}
\right)=\left(
\begin{array}{cc}
1&T_{\mathcal{R}\mathcal{S}}\\
0&T_{\mathcal{S}\mathcal{S}}
\end{array}
\right)\left(\begin{array}{c}
\mathcal{R}\\
\mathcal{S}
\end{array}
\right)_{\ast},
\end{equation}
which gives the power spectra of scalar type perturbations at the
end of inflation
\begin{equation}\label{trans-rel}
\mathcal{P}_{\mathcal{R}}=\mathcal{P}_{\mathcal{R}\ast}+T_{\mathcal{R}\mathcal{S}}^2\mathcal{P}_{\mathcal{S}\ast},~~~~\mathcal{P}_{\mathcal{S}}=T_{\mathcal{S}\mathcal{S}}^2\mathcal{P}_{\mathcal{S}\ast},~~~~\mathcal{P}_{\mathcal{C}}=T_{\mathcal{R}\mathcal{S}}T_{\mathcal{S}\mathcal{S}}\mathcal{P}_{\mathcal{S}\ast}.
\end{equation}
These are the power spectra probed by astronomical observation
\cite{Komatsu:2008hk,:2006uk}, unless they changed significantly
from the end of inflation to the matter-radiation decoupling.

Although the cross-correlation power spectrum is negligibly small
when the perturbations cross the Hubble horizon, it may still be
large at the end of inflation. This is is usually evaluated by the
cross-correlation coefficient $\tilde{\beta}$ \cite{Langlois:1999dw}
(here we use a tilde to distinguish it from the notation appeared in
equation (\ref{evol-uzita})) or the correlation angle $\Delta$
\cite{Wands:2002bn}:
\begin{equation}
\tilde{\beta}=\cos\Delta=\frac{\mathcal{P}_{\mathcal{C}}}{\sqrt{\mathcal{P}_{\mathcal{R}}\mathcal{P}_{\mathcal{S}}}}.
\end{equation}
In terms of the entropy-to-curvature ratio
\begin{equation}\label{encuratio}
r_{\mathcal{S}}=\frac{\mathcal{P}_{\mathcal{S}}}{\mathcal{P}_{\mathcal{R}}},
\end{equation}
it is simplified by the transfer relation (\ref{trans-rel}),
\begin{equation}
\cos\Delta=\sign(T_{\mathcal{S}\mathcal{S}})T_{\mathcal{R}\mathcal{S}}\sqrt{\frac{r_{\mathcal{S}\ast}}{1+T_{\mathcal{R}\mathcal{S}}^2r_{\mathcal{S}\ast}}}=\frac{\sign(T_{\mathcal{S}\mathcal{S}})T_{\mathcal{R}\mathcal{S}}}{\sqrt{1+T_{\mathcal{R}\mathcal{S}}^2}}.
\end{equation}
Here we have used the notation
$\sign(T_{\mathcal{S}\mathcal{S}})=T_{\mathcal{S}\mathcal{S}}/|T_{\mathcal{S}\mathcal{S}}|$
and the fact $r_{\mathcal{S}\ast}\simeq1$. Following the line of
\cite{Byrnes:2006fr}, we expect the coupling terms inside the
horizon would introduce $\mathcal{O}(\sqrt{\epsilon_1})$
corrections to $r_{\mathcal{S}\ast}$.

As a matter of fact, well after the Hubble-exit $k/(aH)\ll1$, the
interaction term $\alpha k^2u_{\mathcal{R} k}$ in equation
(\ref{evol-uds}) is negligible, thus the entropy perturbation
evolves independently outside the horizon. As usually done in
literature, when dealing with the evolution equation of entropy
perturbation, one can ignore the second order term (\emph{i.e.}
$\ddot{\mathcal{S}}_k$ term) and terms proportional to $k^2$.
Recalling relations (\ref{entropy}), (\ref{quant-var}) and
\begin{equation}
u_{\mathcal{S}}=\frac{a\sqrt{F(2F\dot{\varphi}^2+3\dot{F}^2)}}{\sqrt{3}(2HF+\dot{F})}\mathcal{S},
\end{equation}
from equation (\ref{evol-uds}) we can get the evolution equation of
$\mathcal{S}$ to the leading order,
\begin{equation}\label{ex2-transfer1}
\dot{\mathcal{S}}_k=-\frac{H}{3}\left(\frac{m_{\mathcal{S}}^2}{H^2}+2+\epsilon_1-\frac{9}{4}\delta_1-3\delta_2+\frac{3}{2}\delta_4\right)\mathcal{S}_k=\mu_{\mathcal{S}}H\mathcal{S}_k.
\end{equation}
which is a first order differential equation because the second
order term has been ignored. This equation, together with
(\ref{dzeta}) outside the horizon
\begin{eqnarray}\label{ex2-transfer2}
\nonumber \dot{\mathcal{R}}_k&=&\frac{\dot{\varphi}}{\dot{F}}\sqrt{\frac{2F}{3}}\left(\ln\frac{\dot{\varphi}^2}{2F\dot{\varphi}^2+3\dot{F}^2}\right)^{\bullet}\mathcal{S}_k\\
\nonumber &=&\frac{\sign(\dot{\varphi})}{\delta_1}\sqrt{\frac{2(\delta_1-2\epsilon_1)}{3}}\left(2\eta_1-\frac{5}{2}\delta_1-\delta_4\right)H\mathcal{S}_k\\
&=&\mu_{\mathcal{R}}H\mathcal{S}_k,
\end{eqnarray}
dictates the transfer matrix (\ref{trans-matrix}). When evaluating
$m_{\mathcal{S}}^2$ in equation (\ref{ex2-transfer1}), we should keep
the slow-roll order quantities in (\ref{mass-ds}) because
$\mu_{\mathcal{S}}$ is of the slow-roll order. Taking
$\mu_{\mathcal{S}}$ and $\mu_{\mathcal{R}}$ as their average values
between the horizon-exit and the end of inflation, we can quickly
write down the solution for (\ref{ex2-transfer1}) and
(\ref{ex2-transfer2}):
\begin{eqnarray}\label{ex2-sol-trans}
\nonumber &&\mathcal{S}_k=\mathcal{S}_{k\ast}\exp\left(\int_{t_{\ast}}^t\mu_{\mathcal{S}}Hdt\right)=\mathcal{S}_{k\ast}e^{\mu_{\mathcal{S}}(N_{\ast}-N)},\\
&&\mathcal{R}_k-\mathcal{R}_{k\ast}=\int_{t_{\ast}}^t\mu_{\mathcal{R}}H\mathcal{S}_kdt=\int\frac{\mu_{\mathcal{R}}}{\mu_{\mathcal{S}}}\mathcal{S}_{k\ast}de^{-\int\mu_{\mathcal{S}}dN}=\frac{\mu_{\mathcal{R}}}{\mu_{\mathcal{S}}}\mathcal{S}_{k\ast}\left[e^{\mu_{\mathcal{S}}(N_{\ast}-N)}-1\right],
\end{eqnarray}
in which $N=\ln[a_{end}/a(t)]$ stands for the e-folding number from time
$t$ to the end of inflation. Using this solution one may check that
$\ddot{\mathcal{S}}_k/(H\dot{\mathcal{S}}_k)\sim\mathcal{O}(\epsilon_1)$.
So it was reasonable for us to ignore the $\ddot{\mathcal{S}}_k$
term.

With the above results and formulas, it is not hard to obtain the
power spectra at the end of inflation:
\begin{eqnarray}
\nonumber &&\mathcal{P}_{\mathcal{R}}\simeq\mathcal{P}_{\mathcal{R}\ast}+\mathcal{P}_{\mathcal{S}\ast}\frac{\mu_{\mathcal{R}}^2}{\mu_{\mathcal{S}}^2}\left[e^{\mu_{\mathcal{S}}(N_{\ast}-N)}-1\right]^2,\\
\nonumber &&\mathcal{P}_{\mathcal{S}}\simeq\mathcal{P}_{\mathcal{S}\ast}e^{2\mu_{\mathcal{S}}(N_{\ast}-N)},\\
&&\mathcal{P}_{\mathcal{C}}\simeq\mathcal{P}_{\mathcal{S}\ast}\frac{\mu_{\mathcal{R}}}{\mu_{\mathcal{S}}}e^{\mu_{\mathcal{S}}(N_{\ast}-N)}\left[e^{\mu_{\mathcal{S}}(N_{\ast}-N)}-1\right].
\end{eqnarray}
The spectral indices at the end of inflation are
\begin{eqnarray}
\nonumber n_{\mathcal{R}}-1&=&n_{\mathcal{S}\ast}-1-\frac{2\mu_{\mathcal{R}}^2\mu_{\mathcal{S}}e^{\mu_{\mathcal{S}}(N_{\ast}-N)}\left[e^{\mu_{\mathcal{S}}(N_{\ast}-N)}-1\right]}{\mu_{\mathcal{S}}^2+\mu_{\mathcal{R}}^2\left[e^{\mu_{\mathcal{S}}(N_{\ast}-N)}-1\right]^2},\\
\nonumber n_{\mathcal{S}}-1&=&n_{\mathcal{S}\ast}-1-2\mu_{\mathcal{S}},\\
n_{\mathcal{C}}-1&=&n_{\mathcal{S}\ast}-1-\frac{\mu_{\mathcal{S}}\left[2e^{\mu_{\mathcal{S}}(N_{\ast}-N)}-1\right]}{e^{\mu_{\mathcal{S}}(N_{\ast}-N)}-1}.
\end{eqnarray}

In model (\ref{ex2-gV}) with $M_p^2/\varphi^2\ll\lambda
m^2\varphi^2/M_p^2\ll1$, we find $3H\dot{\varphi}>0$, so
$\sign(\dot{\varphi})=1$. Under the slow-roll approximation, we
get
\begin{equation}
\mu_{\mathcal{R}}\simeq-\sqrt{\frac{4\epsilon_1}{3}},~~~~\mu_{\mathcal{S}}\simeq-\frac{29}{3}\epsilon_1,~~~~r_{\mathcal{S}\ast}\simeq1.
\end{equation}
If we estimate $\mu_{\mathcal{R}}$ and $\mu_{\mathcal{R}}$ with
their values at Hubble-exit, the final result will be very simple
(taking $N=0$ at the end of inflation):
\begin{eqnarray}
\nonumber &&\frac{\mathcal{P}_{\mathcal{R}}}{\mathcal{P}_{\mathcal{S}\ast}}\simeq1+\frac{12}{841\epsilon_{1\ast}}\left(e^{-29\epsilon_{1\ast}N_{\ast}/3}-1\right)^2,~~~~\frac{\mathcal{P}_{\mathcal{S}}}{\mathcal{P}_{\mathcal{S}\ast}}\simeq e^{-58\epsilon_{1\ast}N_{\ast}/3},\\
\nonumber &&\frac{\mathcal{P}_{\mathcal{C}}}{\mathcal{P}_{\mathcal{S}\ast}}\simeq\frac{2}{29}\sqrt{\frac{3}{\epsilon_{1\ast}}}e^{-29\epsilon_{1\ast}N_{\ast}/3}\left(e^{-29\epsilon_{1\ast}N_{\ast}/3}-1\right),\\
&&n_{\mathcal{R}}-1=-4\epsilon_{1\ast},~~~~n_{\mathcal{S}}-1=\frac{46}{3}\epsilon_{1\ast},~~~~n_{\mathcal{C}}-1=\frac{17}{3}\epsilon_{1\ast}.
\end{eqnarray}
At the end of inflation, the curvature perturbation and the entropy
perturbation are moderately anti-correlated, with the
cross-correlation coefficient
\begin{equation}
\tilde{\beta}=\cos\Delta\simeq-\sqrt{\frac{12}{841\epsilon_{1\ast}+12}},
\end{equation}
but the entropy-to-curvature ratio is small enough,
\begin{equation}
r_{\mathcal{S}}\simeq\frac{841\epsilon_{1\ast}e^{-58\epsilon_{1\ast}N_{\ast}/3}}{841\epsilon_{1\ast}+12}.
\end{equation}

\begin{figure}
\begin{center}
\includegraphics[width=0.5\textwidth]{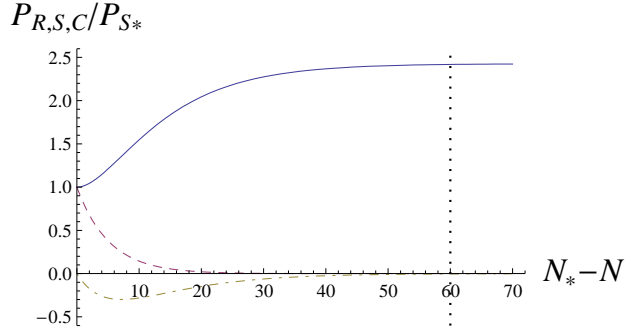}
\end{center}
\caption{\textbf{The evolutions of power spectra with respect to
e-folding number $N_{\ast}-N$ after crossing the horizon. The
curvature power spectrum $\mathcal{P}_{\mathcal{R}}$ is signified by
a blue solid line. The entropy power spectrum
$\mathcal{P}_{\mathcal{S}}$ is denoted by a purple dashed line. The
cross-correlation power spectrum $\mathcal{P}_{\mathcal{C}}$ is
depicted by a brown dot-dashed line. All of the power spectra are
normalized by $\mathcal{P}_{\mathcal{S\ast}}$, the entropy power
spectrum at horizon-crossing. The vertical black dotted line
corresponds to $N_{\ast}-N=60$.}}\label{spectra}
\end{figure}

The evolution of power spectra of curvature and entropy
perturbations and their correlation have been shown in figure
\ref{spectra}. We can see the power spectrum of curvature
perturbation, denoted by the blue solid line, almost doubled from
the horizon-crossing (at $N_{\ast}-N\simeq0$) to the end of inflation
(at $N_{\ast}-N\simeq60$). Although the entropy perturbation (denoted
by the purple dashed line) was lager than the curvature perturbation
at the Hubble-crossing, it was decaying exponentially with respect
to $N_{\ast}-N$ in the super-horizon scale. The correlation between
curvature perturbation and entropy perturbation is depicted by the
brown dot-dashed line. Once crossing out the horizon, it kept a
negative value, with its amplitude at first increasing and then
decreasing.

\begin{figure}
\begin{center}
\includegraphics[width=0.5\textwidth]{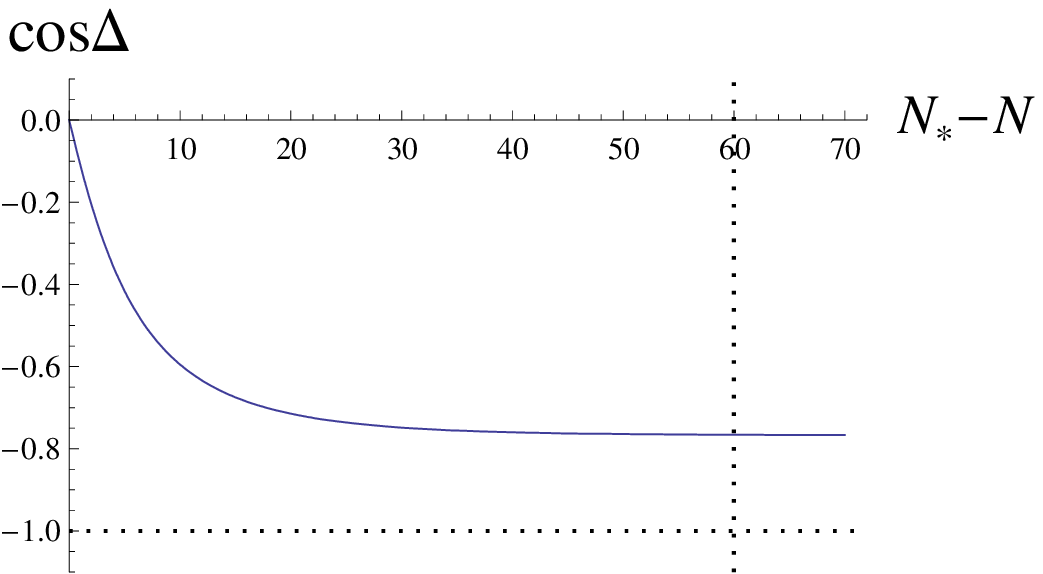}\\
\includegraphics[width=0.5\textwidth]{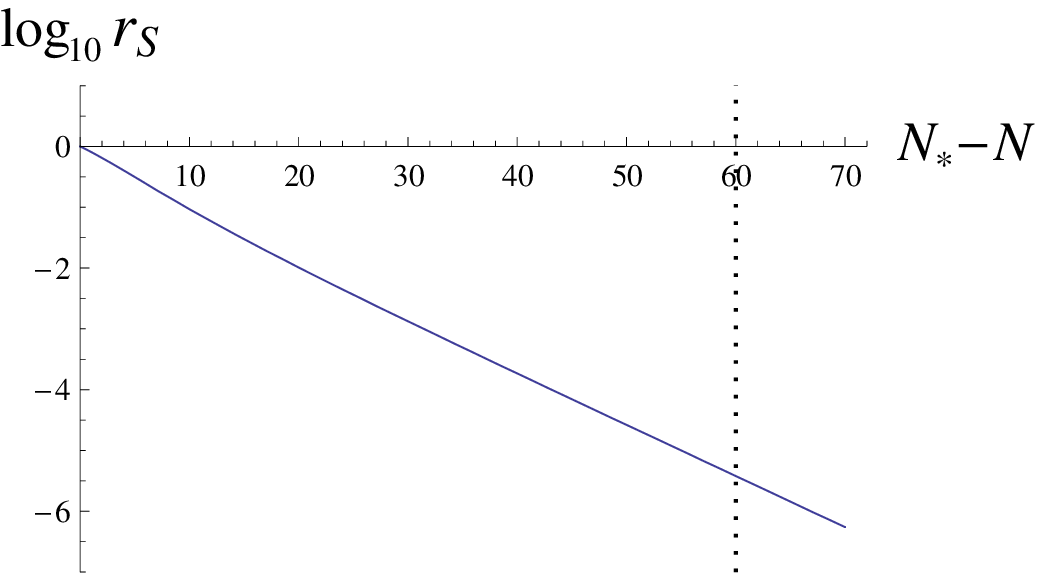}\\
\includegraphics[width=0.5\textwidth]{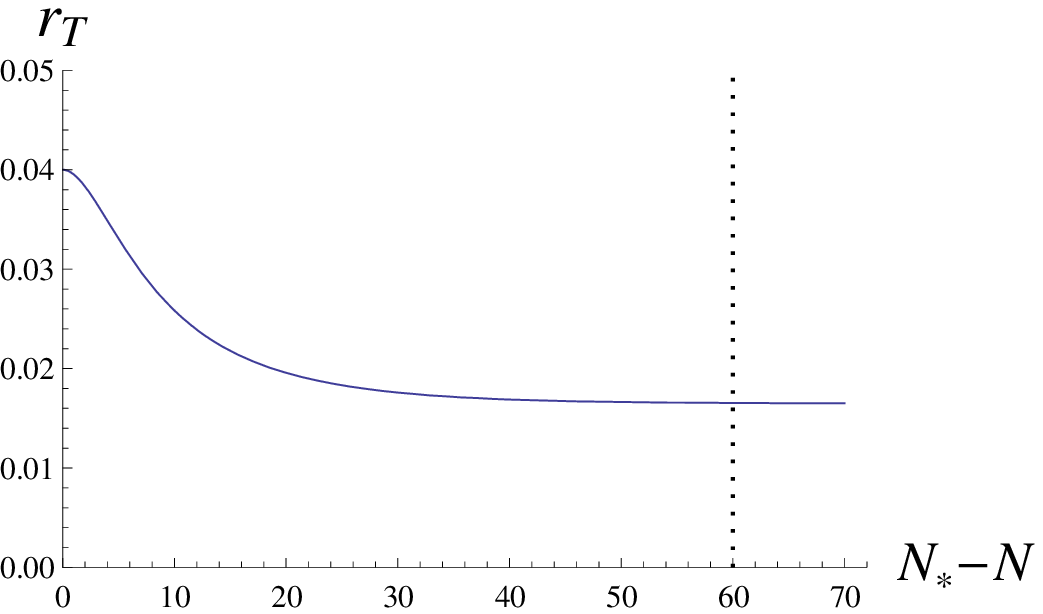}\\
\end{center}
\caption{\textbf{The evolutions of correlation coefficient
$\cos\Delta$ (the top graph), entropy-to-curvature ratio
$r_{\mathcal{S}}$ (its logarithm, the middle graph) and
tensor-to-scalar ratio $r_{T}$ (the bottom graph) with respect to
e-folding number $N_{\ast}-N$ after crossing the horizon. The
vertical black dotted lines correspond to $N_{\ast}-N=60$. The
horizontal black dotted line corresponds to $\cos\Delta=-1$, that
is, the totally anti-correlated situation.}}\label{ratio}
\end{figure}

In figure \ref{ratio}, we illustrate the dependence of correlation
coefficient (the top graph), the logarithm of entropy-to-curvature
ratio (the middle graph) and tensor-to-scalar ratio (the bottom
graph) on $N_{\ast}-N$. It is clear that the entropy-to-curvature
ratio dropped down quickly outside the horizon. The entropy
perturbation and the curvature perturbation was totally
uncorrelated (under the decoupled approximation) at the
Hubble-exit but evolved to be moderately anti-correlated at the
end of inflation.

At the end of inflation, if one assumes
$n_{\mathcal{R}}-1\simeq-0.04$, then
$r_{\mathcal{S}}\simeq4\times10^{-6}$. This is well below the upper
bound put by the five-year Wilkinson Microwave Anisotropy Probe
(WMAP) \cite{Komatsu:2008hk}. Of course, we cannot take these
numbers seriously, because our above analytical results are
meaningful only for estimation. The coupling between entropy
perturbation and curvature perturbation has been systematically
ignored until horizon-crossing. In addition we have estimated
$\mu_{\mathcal{R}}$ and $\mu_{\mathcal{R}}$ with their Hubble-exit
values. Remember that in writing down (\ref{ex2-coe}) we have
neglected the $M_p^4/(2gV)$ term. This may also have introduced some
uncertainties if  $M_p^4/(2gV)\sim\mathcal{O}(\epsilon_1)$. To get a
robust conclusion, one should take all these effects into account
carefully, resorting to the numerical method.

Another issue of our analysis is the special limit we have taken:
$M_p^2/\varphi^2\ll\lambda m^2\varphi^2/M_p^2\ll1$. Taking such a
limit is two-fold. On the one hand, it enables us to get a nearly
scale-invariant entropy perturbation. On the other hand, we find
$3H\dot{\varphi}\simeq m^2\varphi(\lambda
m^2\varphi^4-M_p^4)/M_p^4>0$ in this limit, hence the scalar field
was not rolling down but rolling up the potential $V(\varphi)$.
Also the Hubble parameter is forced to grow ($\epsilon_1>0$). The
same thing also happens in ``phantom'' inflation (single-field
inflation with a kinetic term of the wrong sign), see \emph{e.g.}
\cite{Piao:2007ne}. In our model, as the inflaton $\varphi$ grows,
the slow-roll parameter $\epsilon_1$ tends to order unity and the
inflation is expected to cease then. It is still unclear how to
terminate inflation and trigger reheating in this model. We leave
it as an open problem for future investigation. A more interesting
problem is to get a more realistic, slow-roll model.

According to the result in \cite{Hwang:2005hb}, to the leading
order, the power spectrum of tensor type perturbation in
$f(\varphi,R)$ gravity is
\begin{equation}
\mathcal{P}_{T}\simeq\frac{H^2}{2\pi^2F}=\mathcal{P}_{\mathcal{S}\ast}(2\delta_1-4\epsilon_1).
\end{equation}
It is conserved after Hubble-exit, with a spectral index
\begin{equation}
n_{T}\simeq\frac{2\dot{H}}{H^2}-\frac{\dot{F}}{HF}=2\epsilon_1-\delta_1.
\end{equation}
For the specific model (\ref{ex2-gV}), in the limit
$M_p^2/\varphi^2\ll\lambda m^2\varphi^2/M_p^2\ll1$, they are
\begin{equation}
\frac{\mathcal{P}_{T}}{\mathcal{P}_{\mathcal{S}\ast}}\simeq4\epsilon_{1\ast},~~~~n_{T}\simeq-2\epsilon_{1\ast}.
\end{equation}
In our case, the tensor spectrum is red tilted, which is different
from ``phantom'' inflation. One would have noticed that the
entropy-to-scalar ratio is small at the end of inflation,
\begin{equation}
r_{T}=\frac{\mathcal{P}_{T}}{\mathcal{P}_{\mathcal{R}}}\simeq\frac{3364\epsilon_{1\ast}^2}{841\epsilon_{1\ast}+12}\simeq0.02.
\end{equation}
Its evolution curve outside the horizon is plotted in the lower
graph of figure \ref{ratio}.

\section{Summary}\label{sect-sum}
We investigated the cosmological perturbations in generalized
gravity, where the Ricci scalar and a scalar field are coupled
non-minimally by an arbitrary function $f(\varphi,R)$, with the
Einstein gravity as a special limit. This general form unifies the
usual modified gravity \cite{Sotiriou:2008rp} and scalar-tensor
theory, but often introduces an additional degree of freedom. In the
FLRW background, by studying the first order perturbation theory, we
decomposed the scalar type perturbations into the curvature
perturbation and the entropy perturbation, whose evolution equations
are obtained. The effects of entropy perturbation in this class of
model were seldom studied in the past.

Then we applied this framework to inflation theory. The slow-roll
conditions and the quantized initial condition are discussed. The
quantization of second order action will appear in a future work,
which will put our discussion at the end of section
\ref{sect-slroll} on a firmer ground.

Analytically we studied two special examples: the $\delta s=0$
models in subsection \ref{subsect-noen} and the $g(\varphi)R^2$
model in subsection \ref{subsect-gR2}. In the former example we
unified the previous known results in a unique form and extended
them to the general case. The latter example is more interesting. In
the model $g(\varphi)=\frac{1}{4}\lambda\varphi^2$,
$V(\varphi)=\frac{1}{2}m^2\varphi^2$, we paid attention on the limit
$M_p^2/\varphi^2\ll\lambda m^2\varphi^2/M_p^2\ll1$. This limit helps
us to get a nearly scale-invariant entropy perturbation. Although
the entropy perturbation was large at the Hubble-exit, it decreased
significantly outside the horizon. At the end of inflation, the
``entropy-to-curvature ratio'' we defined by (\ref{encuratio}) was
of order $10^{-6}$, well below the constraint of five-year WMAP
data.

In spite of the above results, there are still some important
problems unsolved. First, the quantization should be performed to
confirm the normalization in (\ref{initialu}) directly. Second, to
get robust conclusions and richer phenomena, maybe one has to employ
a numerical method to solve evolution equations (\ref{evol-uzita})
and (\ref{evol-uds}) more generally. Third, the formalism we
developed in section \ref{sect-curen} can be applied to other
cosmological stages and scenarios, such as the preheating stage and
the curvaton scenario, \emph{etc}. In
\cite{Nojiri:2006ri,Nojiri:2008nt,Nesseris:2008mq,Nesseris:2009jf},
different schemes of modified gravity were investigated mainly
concerning their effects on the late time evolution of our universe,
the possible effect of $f(\varphi,R)$ gravity on late universe is
still waiting for us to study.

\acknowledgments{We are grateful to Bin Chen and Miao Li for reading
a preliminary version of this manuscript. We also thank Yi-Fu Cai,
Xian Gao, Yi Wang, Zhi-Bo Xu and Wei Xue for useful discussions.}

\appendix

\section{Two-field Model and Conformal Transformation}\label{app-conf}
For comparison, let us collect here the well-known results for
two-field inflation with a non-standard kinetic term, which is
conformally equivalent to the $f(\varphi,R)$ generalized gravity
\cite{Teyssandier:1983zz,Maeda:1988ab,Wands:1993uu,Hwang:2005hb}.
Usually the $f(\varphi,R)$ formulism is called Jordan frame while
the non-standard two filed formulism is called Einstein fame. In
this appendix, we use the super/subscripts ``J'' and ``E'' to
distinguish different frames. The action two-field inflation with
a non-standard kinetic term in Einstein fame is
\begin{equation}\label{action-2field}
S=\int d^4x\sqrt{-g_E}\left[\frac{1}{16\pi G}R_E-\frac{1}{2}g_E^{\alpha\beta}\partial^E_{\alpha}\chi\partial^E_{\beta}\chi-\frac{1}{2}e^{2b(\chi)}g_E^{\alpha\beta}\partial^E_{\alpha}\varphi\partial^E_{\beta}\varphi-V(\varphi,\chi)\right].
\end{equation}
The reference \cite{GarciaBellido:1995qq} has an original and
clear discussion on the curvature and entropy perturbations in
this model. The readers may get details there. In a further study
\cite{Gordon:2000hv}, it was clarified that the total entropy
perturbation includes not only the relative entropy perturbation,
but also an entropy perturbation proportional to $k^2/a^2H^2$.
Although it will give a correction at the scale $k\sim aH$, its
contribution to entropy perturbation is suppressed in long
wavelength, so we do not consider its effects in our
investigation.

In longitudinal gauge, the canonical variable in this frame is
given by
\begin{equation}
v_{\chi}=a_E\left(\delta\chi+\frac{\partial^E_t\chi}{H_E}\psi_E\right),~~~~v_{\varphi}=a_Ee^{b}\left(\delta\varphi+\frac{\partial^E_t\varphi}{H_E}\psi_E\right).
\end{equation}
They are related to the canonical variables for curvature and
entropy perturbations
\begin{eqnarray}\label{quant-var2f}
\nonumber v_{\mathcal{R}}&=&\frac{\partial^E_t\chi}{\sqrt{(\partial^E_t\chi)^2+e^{2b}(\partial^E_t\varphi)^2}}v_{\chi}+\frac{e^{b}\partial^E_t\varphi}{\sqrt{(\partial^E_t\chi)^2+e^{2b}(\partial^E_t\varphi)^2}}v_{\varphi}\\
\nonumber &=&\frac{a_E\sqrt{(\partial^E_t\chi)^2+e^{2b}(\partial^E_t\varphi)^2}}{H_E}\mathcal{R},\\
\nonumber v_{\mathcal{S}}&=&\frac{\partial^E_t\chi}{\sqrt{(\partial^E_t\chi)^2+e^{2b}(\partial^E_t\varphi)^2}}v_{\varphi}-\frac{e^{b}\partial^E_t\varphi}{\sqrt{(\partial^E_t\chi)^2+e^{2b}(\partial^E_t\varphi)^2}}v_{\chi}\\
&=&\frac{a_E\sqrt{(\partial^E_t\chi)^2+e^{2b}(\partial^E_t\varphi)^2}}{H_E}\mathcal{S}.
\end{eqnarray}

Making use of the conformal transformation
\begin{equation}
g^E_{\mu\nu}=\frac{F}{M_p^2}g^J_{\mu\nu}
\end{equation}
and the identification
\begin{equation}\label{chi-F}
\frac{\chi}{M_p}=\sqrt{\frac{3}{2}}\ln\frac{F}{M_p^2},
\end{equation}
one can prove the following relations
\cite{Maeda:1988ab,Hwang:1996np,Hwang:2005hb}
\begin{eqnarray}
\nonumber &&a_E=\frac{\sqrt{F}}{M_p}a_J,~~~~dt_E=\frac{\sqrt{F}}{M_p}dt_J,~~~~H_E=\frac{M_p(2H_JF+\partial^J_tF)}{2F^{\frac{3}{2}}},\\
&&\psi_E=\frac{1}{2}(\phi_J+\psi_J),~~~~\delta\chi=\sqrt{\frac{3}{2}}M_p(\psi_J-\phi_J).
\end{eqnarray}

If we take
\begin{equation}
b(\chi)=-\chi/(\sqrt{6}M_p),~~~~V(\phi,\chi)=\frac{M_p^4}{2F^2}[R_JF(\varphi,R_J)-f(\varphi,R_J)],
\end{equation}
then action (\ref{action}) can be perfectly reproduced from action
(\ref{action-2field}). For details of derivation, see reference
\cite{Maeda:1988ab}. In these frames, the conformal time are
coincident $d\tau_E=d\tau_J$, so the canonical quantized variables
(\ref{quant-var2f}) are exactly equivalent to (\ref{quant-var}) by
the above conformal transformation. This conformal equivalence is
powerful. Given a specific function $f(\varphi,R_J)$, one can know
the detailed form of $V(\varphi,\chi)$ using
$F=\frac{\partial}{\partial R_J}f(\varphi,R_J)$ and (\ref{chi-F}),
and then do our job in the more familiar Einstein
frame.\footnote{Thank the referee for an emphasis on this point.}

\section{A Traditional Definition of Curvature and Entropy Perturbations}\label{app-perts}
In all of our discussion, we take the curvature perturbation and
entropy perturbation as defined in (\ref{zeta}) and
(\ref{entrpert}). But such a definition is different from the
traditional one. If we regard the $f(\varphi, R)$ theory as the
Einstein gravity with exotic matter contents induced by the
non-minimal coupling, and following the spirit of
\cite{Bardeen:1980kt}, then we will arrive at a traditional form
of curvature and entropy perturbations. Let us elaborate a little
on this point. One should keep in mind that this point of view is
different from that in appendix \ref{app-conf}. Although we also
use the term ``Einstein gravity'' here, it does not mean the
Einstein frame.

Formally, Friedmann equations (\ref{Friedmann1}) and
(\ref{Friedmann2}) can be rewritten as
\begin{equation}
H^2=\frac{1}{3M_p^{2}}\rho,~~\dot{H}=-\frac{1}{2M_p^{2}}(\rho+p),
\end{equation}
with
\begin{eqnarray}
\nonumber \rho &=&\frac{M_p^{2}}{F}(\frac{1}{2}\dot{\varphi}^2+V+\frac{RF-f}{2}-3H\dot{F}),\\
p&=&\frac{M_p^{2}}{F}(\frac{1}{2}\dot{\varphi}^2-V-\frac{RF-f}{2}+\ddot{F}+2H\dot{F}).
\end{eqnarray}
Here in the effective energy density and pressure we have reckoned
the contribution of the non-minimal coupling. Our final result
seriously depends on this trick. The perturbations obey equations
(\ref{g00}-\ref{gii}). Then the comoving curvature perturbation is
given by
\begin{equation}
\mathcal{R}_{eff}=\psi-\frac{H}{\rho+p}\delta q=\psi-\frac{H}{\dot{H}}\left(\dot{\psi}+H\phi\right).\\
\end{equation}
The curvature perturbation on the uniform density hyper-surface is
well defined,
\begin{equation}
\zeta_{eff}=-\psi-\frac{H}{\dot{\rho}}\delta\rho=-\psi+\frac{H}{\dot{H}}\left(\dot{\psi}+H\phi\right)-\frac{1}{3\dot{H}}\frac{\nabla^2}{a^2}\psi.\\
\end{equation}
At the same time, the entropy perturbation $\delta s_{eff}$ is
defined by
\begin{eqnarray}
\nonumber T\delta s_{eff}&=&\delta p-c_s^2\delta\rho,\\
c_s^2&=&\frac{\partial p}{\partial\rho}=\frac{\dot{p}}{\dot{\rho}}=-\frac{3H\dot{H}+\ddot{H}}{3H\dot{H}}.
\end{eqnarray}
It would be interesting to notice that
\begin{equation}
\zeta_{eff}+\mathcal{R}_{eff}=\frac{2M_p^2}{3(\rho+p)}\frac{\nabla^2}{a^2}\psi.
\end{equation}

Finally, one can quickly prove
\begin{equation}
-\dot{\mathcal{R}}_{eff}=\frac{H}{\dot{H}}\left[\ddot{\psi}+H\dot{\phi}-\frac{\ddot{H}}{\dot{H}}(\dot{\psi}+H\phi)+2\dot{H}\phi\right],
\end{equation}
and
\begin{eqnarray}
\nonumber -\frac{\dot{H}}{H}\dot{\mathcal{R}}_{eff}&=&\frac{1}{2M_p^2}(\delta p-c_s^2\delta\rho)+\frac{c_s^2}{a^2}\nabla^2\psi+\frac{1}{3a^2}\nabla^2(\psi-\phi)\\
&=&\frac{1}{2M_p^2}T\delta s_{eff}+\frac{c_s^2}{a^2}\nabla^2\psi+\frac{1}{3a^2}\nabla^2(\psi-\phi).
\end{eqnarray}

Now we see that the curvature perturbation $\mathcal{R}_{eff}$ and
entropy perturbation $\delta s_{eff}$ are related in the traditional
manner. However, the full expression of $\delta s_{eff}$ is rather
complicated, accordingly its evolution equation is even more
difficult to get. Therefore, the perturbations presented in this
appendix are not convenient in studying the generalized gravity.
Although the definition here appears very natural if one takes
$f(\varphi,R)$ gravity as an effective Einstein theory, a more
convenient choice should be (\ref{zeta}) and (\ref{entrpert}).

\section{Evolution of Entropy Perturbation}\label{app-entr}
From (\ref{zeta}) and (\ref{entrpert}), we get
\begin{equation}\label{dplus}
F(\dot{\phi}+\dot{\psi})=-\frac{1}{2}(2HF+\dot{F})(\phi+\psi)+\frac{2F\dot{\varphi}^2+3\dot{F}^2}{2HF+\dot{F}}\left[\mathcal{R}-\frac{1}{2}(\phi+\psi)\right].
\end{equation}
\begin{equation}\label{minus}
\phi-\psi=\frac{2\dot{F}}{2F\dot{\varphi}^2+3\dot{F}^2}\delta s+\frac{2\dot{F}}{2HF+\dot{F}}\left[\frac{1}{2}(\phi+\psi)-\mathcal{R}\right].
\end{equation}
Now equations (\ref{pert1}) and (\ref{pert2}) can be rewritten in
the form
\begin{eqnarray}\label{ddplus}
\nonumber F(\ddot{\phi}+\ddot{\psi})&=&\left(\frac{2F\ddot{\varphi}}{\dot{\varphi}}-HF-3\dot{F}\right)(\dot{\phi}+\dot{\psi})\\
\nonumber &&+\left[\frac{\ddot{\varphi}}{\dot{\varphi}}(2HF+\dot{F})-(2HF+\dot{F})^{\bullet}\right](\phi+\psi)\\
&&+\left(\frac{3\dot{F}\ddot{\varphi}}{\dot{\varphi}}-\dot{\varphi}^2-3\ddot{F}\right)(\phi-\psi)+\frac{F}{a^2}\nabla^2(\phi+\psi),
\end{eqnarray}
\begin{eqnarray}\label{ddminus}
\nonumber F(\ddot{\phi}-\ddot{\psi})&=&\left(4HF-3\dot{F}+\frac{2F\ddot{\varphi}}{\dot{\varphi}}-\frac{F^2F_{,\varphi}}{3F_{,R}\dot{\varphi}}\right)(\dot{\phi}+\dot{\psi})-3HF(\dot{\phi}-\dot{\psi})\\
\nonumber &&+\left[2F(2H^2+\dot{H})-(2HF+\dot{F})^{\bullet}+(2HF+\dot{F})\left(\frac{\ddot{\varphi}}{\dot{\varphi}}-\frac{FF_{,\varphi}}{6F_{,R}\dot{\varphi}}\right)\right](\phi+\psi)\\
\nonumber &&+\left[2F(2H^2+\dot{H})+\frac{3\dot{F}\ddot{\varphi}}{\dot{\varphi}}-\dot{\varphi}^2-3\ddot{F}-\frac{F^2}{3F_{,R}}-\frac{F\dot{F}F_{,\varphi}}{2F_{,R}\dot{\varphi}}\right](\phi-\psi)\\
&&+\frac{2F}{3a^2}\nabla^2(\phi+\psi)+\frac{F}{a^2}(\phi-\psi).
\end{eqnarray}
Differentiating (\ref{entrpert}) with respect to time and making use
of (\ref{ddplus}), one obtains
\begin{eqnarray}\label{ds}
\nonumber \dot{\delta s}&=&\left(\frac{2F\ddot{\varphi}}{\dot{\varphi}}-\frac{3}{2}\dot{F}\right)(\dot{\phi}+\dot{\psi})+\frac{2F\dot{\varphi}^2+3\dot{F}^2}{2\dot{F}}(\dot{\phi}-\dot{\psi})\\
\nonumber &&+\left[\frac{\ddot{\varphi}}{\dot{\varphi}}(2HF+\dot{F})-\frac{1}{2}(2HF+\dot{F})^{\bullet}\right](\phi+\psi)\\
\nonumber &&+\left[\left(\frac{F\dot{\varphi}^2}{\dot{F}}\right)^{\bullet}+\frac{3\dot{F}\ddot{\varphi}}{\dot{\varphi}}-\dot{\varphi}^2-\frac{3}{2}\ddot{F}\right](\phi-\psi)\\
&&+\frac{F}{a^2}\nabla^2(\phi+\psi),
\end{eqnarray}
which gives
\begin{eqnarray}\label{dminus}
\nonumber &&\frac{2F\dot{\varphi}^2+3\dot{F}^2}{2\dot{F}}(\dot{\phi}-\dot{\psi})\\
\nonumber &=&\dot{\delta s}+\left(\frac{3}{2}\dot{F}-\frac{2F\ddot{\varphi}}{\dot{\varphi}}\right)(\dot{\phi}+\dot{\psi})+\left[\frac{1}{2}(2HF+\dot{F})^{\bullet}-\frac{\ddot{\varphi}}{\dot{\varphi}}(2HF+\dot{F})\right](\phi+\psi)\\
&&+\left[\dot{\varphi}^2+\frac{3}{2}\ddot{F}-\left(\frac{F\dot{\varphi}^2}{\dot{F}}\right)^{\bullet}-\frac{3\dot{F}\ddot{\varphi}}{\dot{\varphi}}\right](\phi-\psi)-\frac{F}{a^2}\nabla^2(\phi+\psi).
\end{eqnarray}
Taking the derivative of (\ref{ds}) once more, we have
\begin{eqnarray}\label{dds}
\nonumber \ddot{\delta s}&=&\left(\frac{2F\ddot{\varphi}}{\dot{\varphi}}-\frac{3}{2}\dot{F}\right)(\ddot{\phi}+\ddot{\psi})+\frac{2F\dot{\varphi}^2+3\dot{F}^2}{2\dot{F}}(\ddot{\phi}-\ddot{\psi})\\
\nonumber &&+\left[\left(\frac{2F\ddot{\varphi}}{\dot{\varphi}}\right)^{\bullet}+\frac{\ddot{\varphi}}{\dot{\varphi}}(2HF+\dot{F})-(HF+2\dot{F})^{\bullet}\right](\dot{\phi}+\dot{\psi})\\
\nonumber &&+\left[\left(\frac{2F\dot{\varphi}^2}{\dot{F}}\right)^{\bullet}+\frac{3\dot{F}\ddot{\varphi}}{\dot{\varphi}}-\dot{\varphi}^2\right](\dot{\phi}-\dot{\psi})\\
\nonumber &&+\left\{\left[\frac{\ddot{\varphi}}{\dot{\varphi}}(2HF+\dot{F})\right]^{\bullet}-\frac{1}{2}(2HF+\dot{F})^{\bullet\bullet}\right\}(\phi+\psi)\\
\nonumber &&+\left[\left(\frac{F\dot{\varphi}^2}{\dot{F}}\right)^{\bullet\bullet}+\left(\frac{3\dot{F}\ddot{\varphi}}{\dot{\varphi}}-\dot{\varphi}^2-\frac{3}{2}\ddot{F}\right)^{\bullet}\right](\phi-\psi)\\
&&+\frac{F}{a^2}\nabla^2(\dot{\phi}+\dot{\psi})+(\dot{F}-2HF)\frac{\nabla^2}{a^2}(\phi+\psi).
\end{eqnarray}
Substituting (\ref{ddplus}) and (\ref{ddminus}), (\ref{dminus}),
(\ref{dplus}) and (\ref{minus}) into (\ref{dds}) step by step to
eliminate $\ddot{\phi}+\ddot{\psi}$ and  $\ddot{\phi}-\ddot{\psi}$,
$\dot{\phi}-\dot{\psi}$, $\dot{\phi}+\dot{\psi}$ and $\phi-\psi$
respectively, one will arrive at a rather scattering equation:
\begin{eqnarray}
\nonumber \ddot{\delta s}&=&\left\{\left[\ln\frac{\dot{\varphi}^2(2F\dot{\varphi}^2+3\dot{F}^2)}{\dot{F}^2}\right]^{\bullet}-3H\right\}\dot{\delta s}+\frac{\nabla^2}{a^2}\delta s\\
\nonumber &&+\Biggl\{\left[\frac{\ddot{\varphi}}{\dot{\varphi}}(2HF+\dot{F})\right]^{\bullet}-\frac{1}{2}(2HF+\dot{F})^{\bullet\bullet}+\left(\frac{2\ddot{\varphi}}{\dot{\varphi}}-\frac{3\dot{F}}{2F}\right)\left[\frac{\ddot{\varphi}}{\dot{\varphi}}(2HF+\dot{F})-(2HF+\dot{F})^{\bullet}\right]\\
\nonumber &&+\left(\frac{\dot{\varphi}^2}{\dot{F}}+\frac{3\dot{F}}{2F}\right)\left[2F(2H^2+\dot{H})-(2HF+\dot{F})^{\bullet}+(2HF+\dot{F})\left(\frac{\ddot{\varphi}}{\dot{\varphi}}-\frac{FF_{,\varphi}}{6F_{,R}\dot{\varphi}}\right)\right]\\
\nonumber &&+\left[\left(\ln\frac{\dot{\varphi}^2(2F\dot{\varphi}^2+3\dot{F}^2)}{\dot{F}^2}\right)^{\bullet}-3H\right]\left[\frac{1}{2}(2HF+\dot{F})^{\bullet}-\frac{\ddot{\varphi}}{\dot{\varphi}}(2HF+\dot{F})\right]\Biggr\}(\phi+\psi)\\
\nonumber &&+\Biggl\{\left(\frac{2F\ddot{\varphi}}{\dot{\varphi}}\right)^{\bullet}+\frac{\ddot{\varphi}}{\dot{\varphi}}(2HF+\dot{F})-(HF+2\dot{F})^{\bullet}+\left(\frac{2\ddot{\varphi}}{\dot{\varphi}}-\frac{3\dot{F}}{2F}\right)\left(\frac{2F\ddot{\varphi}}{\dot{\varphi}}-HF-3\dot{F}\right)\\
\nonumber &&+\left(\frac{\dot{\varphi}^2}{\dot{F}}+\frac{3\dot{F}}{2F}\right)\left(4HF-3\dot{F}+\frac{2F\ddot{\varphi}}{\dot{\varphi}}-\frac{F^2F_{,\varphi}}{3F_{,R}\dot{\varphi}}\right)\\
\nonumber &&+\left[\left(\ln\frac{\dot{\varphi}^2(2F\dot{\varphi}^2+3\dot{F}^2)}{\dot{F}^2}\right)^{\bullet}-3H\right]\left(\frac{3}{2}\dot{F}-\frac{2F\ddot{\varphi}}{\dot{\varphi}}\right)\Biggr\}\\
\nonumber &&\times\Biggl\{-\frac{2HF+\dot{F}}{2F}(\phi+\psi)-\left(\frac{\dot{\varphi}^2}{\dot{F}}+\frac{3\dot{F}}{2F}\right)\frac{2\dot{F}}{2HF+\dot{F}}\left[\frac{1}{2}(\phi+\psi)-\mathcal{R}\right]\Biggr\}\\
\nonumber &&+\Biggl\{\left(\frac{F\dot{\varphi}^2}{\dot{F}}\right)^{\bullet\bullet}+\left(\frac{3\dot{F}\ddot{\varphi}}{\dot{\varphi}}-\dot{\varphi}^2-\frac{3}{2}\ddot{F}\right)^{\bullet}+\left(\frac{2\ddot{\varphi}}{\dot{\varphi}}-\frac{3\dot{F}}{2F}\right)\left(\frac{3\dot{F}\ddot{\varphi}}{\dot{\varphi}}-\dot{\varphi}^2-3\ddot{F}\right)\\
\nonumber &&+\left(\frac{\dot{\varphi}^2}{\dot{F}}+\frac{3\dot{F}}{2F}\right)\left[2F(2H^2+\dot{H})+\frac{3\dot{F}\ddot{\varphi}}{\dot{\varphi}}-\dot{\varphi}^2-3\ddot{F}-\frac{F^2}{3F_{,R}}-\frac{F\dot{F}F_{,\varphi}}{2F_{,R}\dot{\varphi}}\right]\\
\nonumber &&+\left[\left(\ln\frac{\dot{\varphi}^2(2F\dot{\varphi}^2+3\dot{F}^2)}{\dot{F}^2}\right)^{\bullet}-3H\right]\left[\dot{\varphi}^2+\frac{3}{2}\ddot{F}-\left(\frac{F\dot{\varphi}^2}{\dot{F}}\right)^{\bullet}-\frac{3\dot{F}\ddot{\varphi}}{\dot{\varphi}}\right]\Biggr\}\\
\nonumber &&\times\Biggl\{\frac{2\dot{F}}{2F\dot{\varphi}^2+3\dot{F}^2}\delta s+\frac{2\dot{F}}{2HF+\dot{F}}\left[\frac{1}{2}(\phi+\psi)-\mathcal{R}\right]\Biggr\}\\
\nonumber &&+\Biggl\{-\frac{1}{2}(2HF+\dot{F})+\dot{F}-2HF+\frac{2F\ddot{\varphi}}{\dot{\varphi}}-\frac{3}{2}\dot{F}+\frac{2F\dot{\varphi}^2}{3\dot{F}}+\dot{F}\\
&&-F\left[\ln\frac{\dot{\varphi}^2(2F\dot{\varphi}^2+3\dot{F}^2)}{\dot{F}^2}\right]^{\bullet}+3HF\Biggr\}\frac{\nabla^2}{a^2}(\phi+\psi).
\end{eqnarray}
This equation looks terribly lengthy. However, repeatedly employing
the relation (\ref{Friedmann2}) or namely
\begin{equation}
(2HF+\dot{F})^{\bullet}=3H\dot{F}-\dot{\varphi}^2,
\end{equation}
after careful calculation, we find the coefficients of the
$\frac{2\dot{F}}{2HF+\dot{F}}\left[\frac{1}{2}(\phi+\psi)-\mathcal{R}\right]$
and remaining $(\phi+\psi)$ terms are exactly vanished, resulting in
a much simpler form (\ref{evol-s}).


\begin{thebibliography}{99}

\bibitem{Guth:1980zm}
  A.~H.~Guth,
  Phys.\ Rev.\  D {\bf 23}, 347 (1981).

\bibitem{Linde:1981mu}
  A.~D.~Linde,
  Phys.\ Lett.\  B {\bf 108}, 389 (1982).

\bibitem{Albrecht:1982wi}
  A.~Albrecht and P.~J.~Steinhardt,
  Phys.\ Rev.\ Lett.\  {\bf 48}, 1220 (1982).

\bibitem{AdelmanMcCarthy:2007wh}
  J.~K.~Adelman-McCarthy  [SDSS Collaboration],
  Astrophys.\ J.\ Suppl.\  {\bf 172}, 634 (2007)
  [arXiv:0707.3380 [astro-ph]].

\bibitem{Komatsu:2008hk}
  E.~Komatsu {\it et al.}  [WMAP Collaboration],
  Astrophys.\ J.\ Suppl.\  {\bf 180}, 330 (2009)
  [arXiv:0803.0547 [astro-ph]].

\bibitem{:2006uk}
    [Planck Collaboration],
  arXiv:astro-ph/0604069.

\bibitem{Mukhanov:1990me}
  V.~F.~Mukhanov, H.~A.~Feldman and R.~H.~Brandenberger,
  Phys.\ Rept.\  {\bf 215}, 203 (1992).

\bibitem{Riotto:2002yw}
  A.~Riotto,
  arXiv:hep-ph/0210162.

\bibitem{Polarski:1992dq}
  D.~Polarski and A.~A.~Starobinsky,
  Nucl.\ Phys.\  B {\bf 385}, 623 (1992).

\bibitem{Polarski:1994rz}
  D.~Polarski and A.~A.~Starobinsky,
  Phys.\ Rev.\  D {\bf 50}, 6123 (1994)
  [arXiv:astro-ph/9404061].

\bibitem{Polarski:1995zn}
  D.~Polarski and A.~A.~Starobinsky,
  Phys.\ Lett.\  B {\bf 356}, 196 (1995)
  [arXiv:astro-ph/9505125].

\bibitem{Langlois:1999dw}
  D.~Langlois,
  Phys.\ Rev.\  D {\bf 59}, 123512 (1999)
  [arXiv:astro-ph/9906080].

\bibitem{Gordon:2000hv}
  C.~Gordon, D.~Wands, B.~A.~Bassett and R.~Maartens,
  Phys.\ Rev.\  D {\bf 63}, 023506 (2001)
  [arXiv:astro-ph/0009131].

\bibitem{Wands:2002bn}
  D.~Wands, N.~Bartolo, S.~Matarrese and A.~Riotto,
  Phys.\ Rev.\  D {\bf 66}, 043520 (2002)
  [arXiv:astro-ph/0205253].

\bibitem{Starobinsky:1980te}
  A.~A.~Starobinsky,
  Phys.\ Lett.\  B {\bf 91}, 99 (1980).

\bibitem{Starobinsky:1983sov}
  A.~A.~Starobinsky,
  Sov. Astron. Lett. 9 (1983) 302.

\bibitem{Teyssandier:1983zz}
  P.~Teyssandier and Ph.~Tourrenc,
  J.\ Math.\ Phys.\  {\bf 24}, 2793 (1983).

\bibitem{Maeda:1988ab}
  K.~I.~Maeda,
  Phys.\ Rev.\  D {\bf 39}, 3159 (1989).

\bibitem{Wands:1993uu}
  D.~Wands,
  Class.\ Quant.\ Grav.\  {\bf 11}, 269 (1994)
  [arXiv:gr-qc/9307034].

\bibitem{Hwang:1990re}
  J.~C.~Hwang,
  Class.\ Quant.\ Grav.\  {\bf 7}, 1613 (1990).

\bibitem{Hwang:1996np}
  J.~C.~Hwang,
  Class.\ Quant.\ Grav.\  {\bf 14}, 1981 (1997)
  [arXiv:gr-qc/9605024].

\bibitem{Hwang:1996bc}
  J.~C.~Hwang,
  Class.\ Quant.\ Grav.\  {\bf 14}, 3327 (1997)
  [arXiv:gr-qc/9607059].

\bibitem{Hwang:1997uc}
  J.~C.~Hwang,
  Class.\ Quant.\ Grav.\  {\bf 15}, 1401 (1998)
  [arXiv:gr-qc/9710061].

\bibitem{Hwang:2005hb}
  J.~C.~Hwang and H.~Noh,
  Phys.\ Rev.\  D {\bf 71}, 063536 (2005)
  [arXiv:gr-qc/0412126].

\bibitem{Chen:2006wn}
  B.~Chen, M.~Li, T.~Wang and Y.~Wang,
  Mod.\ Phys.\ Lett.\  A {\bf 22}, 1987 (2007)
  [arXiv:astro-ph/0610514].

\bibitem{Lyth:2001nq}
  D.~H.~Lyth and D.~Wands,
  Phys.\ Lett.\  B {\bf 524}, 5 (2002)
  [arXiv:hep-ph/0110002].

\bibitem{Lyth:2002my}
  D.~H.~Lyth, C.~Ungarelli and D.~Wands,
  Phys.\ Rev.\  D {\bf 67}, 023503 (2003)
  [arXiv:astro-ph/0208055].

\bibitem{Huang:2008ze}
  Q.~G.~Huang,
  Phys.\ Lett.\  B {\bf 669}, 260 (2008)
  [arXiv:0801.0467 [hep-th]].

\bibitem{Huang:2008rj}
  Q.~G.~Huang,
  JCAP {\bf 0809}, 017 (2008)
  [arXiv:0807.1567 [hep-th]].

\bibitem{Huang:2008bg}
  Q.~G.~Huang and Y.~Wang,
  JCAP {\bf 0809}, 025 (2008)
  [arXiv:0808.1168 [hep-th]].

\bibitem{GarciaBellido:1995qq}
  J.~Garcia-Bellido and D.~Wands,
  Phys.\ Rev.\  D {\bf 53}, 5437 (1996)
  [arXiv:astro-ph/9511029].

\bibitem{GrootNibbelink:2001qt}
  S.~Groot Nibbelink and B.~J.~W.~van Tent,
  Class.\ Quant.\ Grav.\  {\bf 19}, 613 (2002)
  [arXiv:hep-ph/0107272].

\bibitem{DiMarco:2002eb}
  F.~Di Marco, F.~Finelli and R.~Brandenberger,
  Phys.\ Rev.\  D {\bf 67}, 063512 (2003)
  [arXiv:astro-ph/0211276].

\bibitem{DiMarco:2005nq}
  F.~Di Marco and F.~Finelli,
  Phys.\ Rev.\  D {\bf 71}, 123502 (2005)
  [arXiv:astro-ph/0505198].

\bibitem{Bardeen:1980kt}
  J.~M.~Bardeen,
  Phys.\ Rev.\  D {\bf 22}, 1882 (1980).

\bibitem{Li:2006te}
  M.~Li,
  JCAP {\bf 0610}, 003 (2006)
  [arXiv:astro-ph/0607525].

\bibitem{Byrnes:2006fr}
  C.~T.~Byrnes and D.~Wands,
  Phys.\ Rev.\  D {\bf 74}, 043529 (2006)
  [arXiv:astro-ph/0605679].

\bibitem{Piao:2007ne}
  Y.~S.~Piao,
  Phys.\ Rev.\  D {\bf 78}, 023518 (2008)
  [arXiv:0712.3328 [gr-qc]].

\bibitem{Sotiriou:2008rp}
  T.~P.~Sotiriou and V.~Faraoni,
  arXiv:0805.1726 [gr-qc].

\bibitem{Nojiri:2006ri}
  S.~Nojiri and S.~D.~Odintsov,
  eConf {\bf C0602061}, 06 (2006)
  [Int.\ J.\ Geom.\ Meth.\ Mod.\ Phys.\  {\bf 4}, 115 (2007)]
  [arXiv:hep-th/0601213].

\bibitem{Nojiri:2008nt}
  S.~Nojiri and S.~D.~Odintsov,
  arXiv:0807.0685 [hep-th].

\bibitem{Nesseris:2008mq}
  S.~Nesseris,
  arXiv:0811.4292 [astro-ph].

\bibitem{Nesseris:2009jf}
  S.~Nesseris and A.~Mazumdar,
  arXiv:0902.1185 [astro-ph.CO].
\end{thebibliography}
\end{document}